\documentclass{aa}
\usepackage{graphicx}
\usepackage[]{txfonts}
\newcommand{\teff}{T_{\rm eff}}

\newcommand{\Ha}{H$\alpha$}
\newcommand{\FeI}{Fe\,{\sc i}\,}
\newcommand{\FeII}{Fe\,{\sc ii}}

\newcommand{\BaII}{Ba\,{\sc ii}\,}
\newcommand{\CaI}{Ca\,{\sc i}\,}
\newcommand{\CaII}{Ca\,{\sc ii}\,}

\newcommand{\SiII}{Si\,{\sc ii}\,}
\newcommand{\LiI}{Li\,{\sc i}\,}
\newcommand{\VI}{V\,{\sc i}\,}
\newcommand{\ScII}{Sc\,{\sc ii}\,}
\newcommand{\LaII}{La\,{\sc ii}\,}
\newcommand{\YII}{Y\,{\sc ii}\,}

\newcommand{\kms}{km\,s$^{-1}$}

\begin{document}
   \title{Identification of a complete sample of northern ROSAT All-Sky \newline Survey X-ray sources
}
     \subtitle{
     VIII. The late-type stellar component
     \thanks{Based on observations collected at the German-Spanish Astronomical Centre, Calar Alto, 
operated by the Max-Planck-Institut f\"ur Astronomie, Heidelberg,
jointly with the Spanish National Commission for Astronomy, and at the European Southern 
Observatory, La Silla, Chile.}
}
\author{F.-J.~Zickgraf \inst{1} \and J. Krautter \inst{2} \and  S. Reffert \inst{3} 
\and J.M. Alcal\'a \inst{4} \and R.
Mujica \inst{5}  \and E. Covino,\inst{4}  \and M.F. Sterzik \inst{6}}
  
\institute{Hamburger Sternwarte, Gojenbergsweg 112, 21029 Hamburg, Germany
\and Landessternwarte K\"onigstuhl, D-69117 Heidelberg, Germany
\and Sterrewacht Leiden, P.O. Box 9513, NL-2300 RA Leiden, The Netherlands
\and Osservatorio Astronomico di Capodimonte, Via Moiariello 16, I-80131 Napoli,
Italy
\and Instituto Nacional de Astrofisica, Optica y Electronica, 
A. Postal 51 y 216 Z.P., 72000 Puebla, Mexico
\and European Southern Observatory, Alonso de Cordova 3107, Santiago 19, Chile
}
\offprints{F.-J. Zickgraf}
\date{Received 16 August 2004/ Accepted 03 December 2004}
\abstract{
We present results of an investigation of the X-ray properties, age distribution, and
kinematical characteristics of a high-galactic 
latitude sample of late-type field stars selected from 
the ROSAT All-Sky Survey (RASS). The sample comprises 254 RASS sources
with optical counterparts of spectral types F to M distributed over six study areas 
located at $|b| \ga 20\degr$, and $DEC \ge -9\degr$. A detailed study was carried out for 
the subsample of 
$\sim$200  G, K, and M stars. Lithium abundances 
were determined for 179 G-M stars. Radial velocities were measured for 
most of the 141 G and K type stars of the sample. 
Combined with proper motions these data were used to study the age 
distribution and the kinematical properties of the sample. 
Based on the lithium abundances half of the G-K stars were 
found to be younger than the Hyades (660\,Myr). About 25\% are 
comparable in age to the Pleiades (100\,Myr). A small 
subsample of 10 stars is younger than the Pleiades. 
They are therefore most likely pre-main sequence stars. 
Kinematically the PMS and Pleiades-type stars appear to form a group 
with space velocities close to the Castor moving group but 
clearly distinct from the Local Association. 

}

\titlerunning{The late-type stellar component in the RASS at high $|b|$}
\authorrunning{F.-J. Zickgraf et al.}
\maketitle

\section{Introduction}
\label{intro}
In a series of previous papers we reported about the results of a large programme on the 
optical identification of a complete 
count-rate limited
sample of northern high-galactic latitude X-ray 
sources from the ROSAT All-Sky Survey (RASS)
(Zickgraf et al. \cite{pap2}, \cite{pap5}, \cite{pap6}, Appenzeller 
et al. \cite{pap3}, \cite{pap7}, Krautter et al. \cite{pap4}). 
The sample was selected for the purpose of the investigation of the
statistical composition of the high-galactic latitude part of the RASS 
in the northern hemisphere.
As described in detail by Zickgraf et al. (\cite{pap2}) (hereafter Paper II) 
the selection criteria for the X-ray sources 
were  X-ray count-rate and location in the sky.
The sample is 
distributed in six study areas located at 
galactic latitudes 
$|b| \ge 20\degr$ and north of declination $-9\degr$. The optical identification 
was based on multi-object spectroscopy  and direct CCD imaging. 
For 
more information on the identification process cf. Paper II.
The catalogue of optical identifications and the statistical analysis of the sample
were presented in Appenzeller et al. (\cite{pap3}) and Krautter et al. (\cite{pap4})
(hereafter Paper III and IV, respectively).
We found that about 60\% of the selected X-ray sources are extragalactic
objects, i.e. AGN, clusters of galaxies, and individual galaxies. 
About 40\% are stellar 
sources. Most of these (257 out of 274 objects) are F-M type coronal emitters. The rest are
cataclysmic variables and white dwarfs. 
Follow-up investigations on the properties of subsamples formed by certain object classes 
were carried out for AGN (Appenzeller et al. \cite{pap7}) and galaxy clusters 
(Appenzeller et al. \cite{App00}).
A paper on the BL Lac objects in the sample is in preparation (Mujica et al.). 
The paper presented here is dedicated to the characteristics of the coronal
stellar component.

A first discussion of the properties of  the late-type stellar component was given 
by Zickgraf et al. (\cite{pap6}) (Paper VI). A subsample of stars in study area\,I 
located some $20\degr$ south of the Taurus-Auriga star forming region 
(SFR) was found to contain a large fraction of very young, presumably pre-main 
sequence stars. In order to investigate the age distribution of the complete 
sample of coronal 
X-ray emitters we obtained further low, medium and/or high resolution spectroscopic observations 
for most G-M stars in our sample. 
The goals were to carry out a lithium survey in order to identify lithium-rich high-galactic 
latitude G-M type stars and to determine precise radial velocities. 
In solar-like stars the  lithium abundance can be used as an age estimator. Its knowledge
therefore allows to study the age distribution of the X-ray active stellar sample. Combining the 
age information with proper motions and radial velocities would thus allow to investigate a possible 
age dependence of the kinematical properties of the stellar RASS sample.      

This paper is structured as follows. The sample is presented in Sect. \ref{sample}. 
Observations and data reduction are described in Sect. \ref{obs}. 
Observational results are presented in Sect. \ref{obsresult}. 
Based on these results the sample properties are analysed and 
discussed in Sect. \ref{analysis}. 
Finally, conclusions are given in Sect. \ref{sum}.

\begin{table*} 
\centering
\caption[]{Journal of observations.} 
\begin{tabular}{lcll} 
\noalign{\smallskip}     
\hline 
\hline 
\noalign{\smallskip}    
date                        &  spectral res.                 & instrument & telescope\\
                            &  $R = \lambda / \Delta\lambda$ &            &            \\ 
\hline 
\noalign{\smallskip}    
Sep. 20-23, 1996           & 2\,100  & CAFOS   &CA 2.2\,m \\
Jan. 31 - Feb. 3,  1997   & 2\,100  & CAFOS   &CA 2.2\,m \\
Jan. 2-6, 1998             & 1\,600  & CAFOS   &CA 2.2\,m \\
Feb. 18, 1998              & 22\,000 & CASPEC  & ESO 3.6\,m\\
May 14-16,  1998           & 1\,300  & DFOSC   & ESO Danish 1.54\,m\\
Dec. 22-25, 1998           & 34\,000 & FOCES   & CA 2.2\,m\\
Apr. 29 - May 4, 1998      & 4\,600  & CARELEC & OHP 1.93\,m\\
Oct. 21-16, 1998           & 20\,000 & AURELIE & OHP 1.52\,m\\
Jan. 11-15, 2000           & 34\,000 & FOCES   & CA 2.2\,m\\
Jun. 13-18, 2000           & 34\,000 & FOCES   & CA 2.2\,m\\
Dec. 3-6, 2001             & 34\,000 & FOCES   & CA 2.2\,m\\
Feb. 19-23, 2002           & 34\,000 & FOCES   & CA 2.2\,m\\
\hline 
\noalign{\smallskip}    
\end{tabular}
\label{journal}
\end{table*}

\section{The  sample}
\label{sample}
\subsection{Sample selection}
\begin{figure*}[tbh]
\resizebox{\hsize}{!}{\includegraphics[angle=-90,bbllx=80pt,bblly=60pt,bburx=540pt,bbury=770pt,clip=]{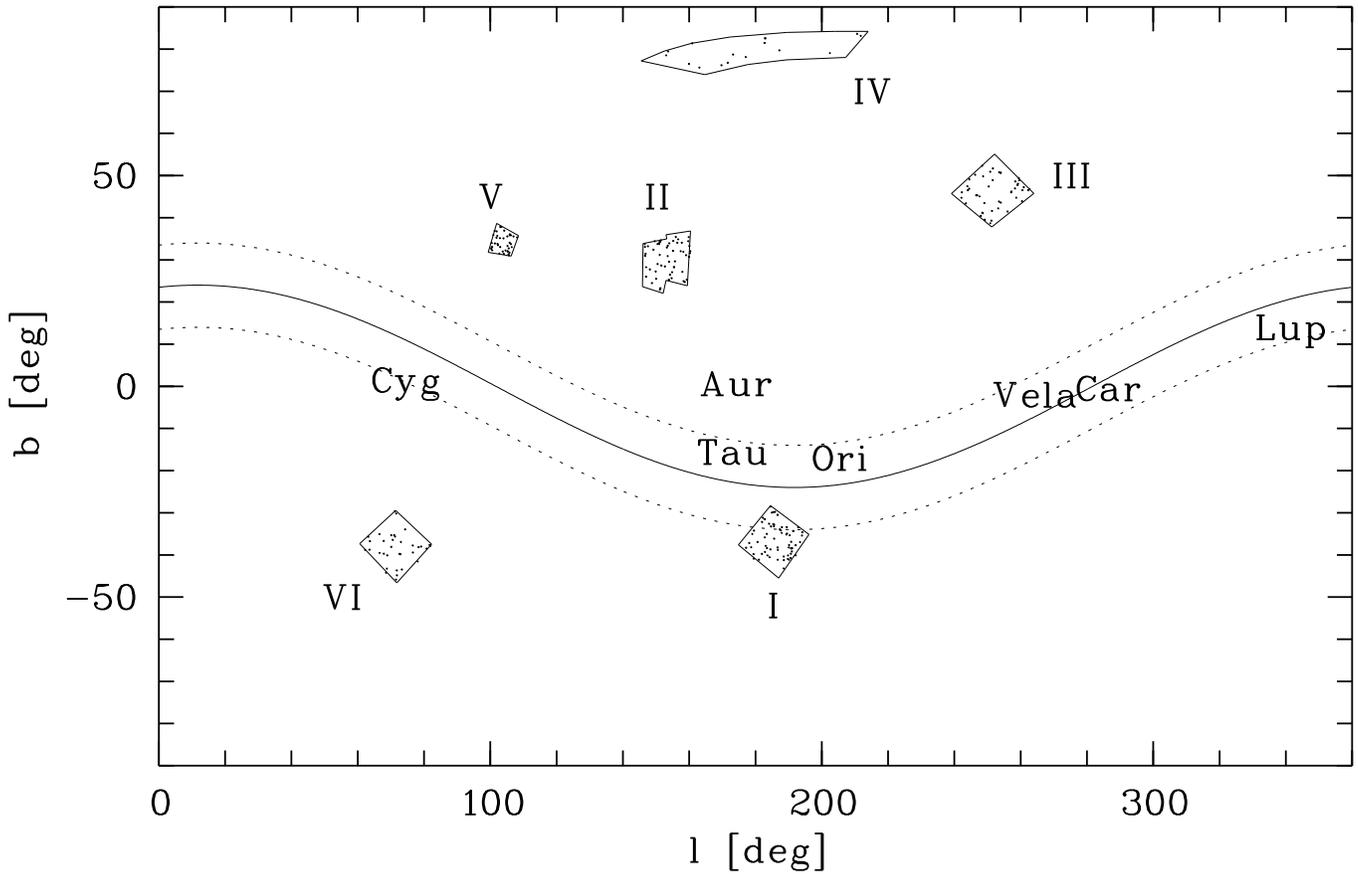}}
\caption[]{ Location of the six study areas in galactic coordinates. The dots show the positions
of the RASS X-ray sources with stellar counterparts in the respective area. The solid and dashed curves
denote the position and width, respectively, of the Gould Belt according to Guillout et al. 
(\cite{Guilloutetal98}). 
In addition positions of several associations are shown.
}
\label{area_gal}
\end{figure*}

Paper III presents a catalogue of optical identifications for 685 RASS sources contained 
in six study areas. The location of the study areas is plotted  in Fig. \ref{area_gal} 
in galactic coordinates. The catalogue contains 254 X-ray sources which have been 
identified as coronal emitters of spectral types F to M. The contribution of the different
spectral types is given in Table \ref{revstat}. One X-ray source, 
E020 = \object{RXJ\,1627.8+7042}, was dropped from the stellar subsample. The star assigned as 
counterpart to this RASS source is too far 
from the X-ray position to be a plausible identification ($d =$ 1.5 arcmin). This source 
is more likely an optically faint AGN.

For the spectroscopic follow-up investigation we selected
the 200 X-ray sources from the catalogue with stellar counterparts of 
spectral types G to M. F stars were not included in the spectroscopic follow-up 
observations because for these stars the lithium abundance is not a good age 
estimator. In 19 cases two stars have been assigned as counterpart to the X-ray source 
in Paper III. Several of these secondary counterparts were also observed. As 
in Paper IV we will however only use the primary identifications for statistical 
purposes. The entire
``coronal'' sample including the F type stars comprises 253 X-ray sources. 
The known RS CVn star \object{HR\,1099} (= \object{V\,711\,Tau}) which is  X-ray source A031 
in Paper III was excluded from the coronal sample
discussed in the following. The sample finally selected for spectroscopic 
follow-up observations thus comprised 199 of the 200 X-ray sources with optical 
counterparts of spectral type G to M as listed in Paper III. 
 
\subsection{Photometry}
Paper III gives visual magnitudes based mainly on photographic 
photo\-metry from the Automated Plate Measuring (APM) machine  (Irwin \& McMahon \cite{IrwinMcMahon92})
except for stars with photometry listed in SIMBAD. Because the APM magnitudes
lack a proper calibration their accuracy is rather low. We therefore obtained 
improved $V$ magnitudes by using photometry from other sources. Bright stars are contained in 
the Tycho-2 catalogue (Hog et al. \cite{Hogetal00}). For fainter stars  not found in Tycho-2 
magnitudes were taken either from the Hubble Guide Star Catalogue GSC-I (Lasker et al.
\cite{Laskeretal90}) or for stars fainter than the limit of  GSC-I  
from the GSC-II catalogue (The Guide Star Catalogue, Version 2.2.01). 
In a few cases no photometry is available because of blending 
with nearby neighbours. The Tycho-2 $V_T$ magnitudes were transformed to  Johnson $V$
according to Mamajek et al. (\cite{Mamajeketal02}). GSC-I magnitudes were transformed 
to Johnson $V$ using the colour coefficients given in Russell et al. 
(\cite{Russelletal90}) and $B-V$ colours 
of main sequence stars for the corresponding spectral type taken from Schmidt-Kaler 
(\cite{SchmidtKaler82}). These colours were also used to calculate Johnson $V$ 
magnitudes from the GSC-II $B$ magnitudes. The improved $V$ magnitudes were then 
used to recalculate the ratio of X-ray-to-optical flux, $f_{\rm x}/f_V$, which is 
given in the Appendix in  Table \ref{basic} 
together with other basic parameters of the sample stars.

Infrared photometry in $J, H,$ and $K$ was taken from the Two Micron All Sky Survey 
(2MASS) catalogue. From this data base infrared sources within 10\arcsec\ around the optical 
position of the counterpart were extracted. A total of 267 2MASS sources was found of which
90\% were located within 2\arcsec\ from the optical counterparts (including the 19 double
identifications, see above). We  considered the 258 matches within 4\arcsec , i.e. within
$3\sigma$ as reliable identifications. Matches between 4\arcsec\  and 10\arcsec\ were 
individually checked and all found to be also correct. This means that for all but 5 RASS 
sources  (A035, A045, A065, D022, and D114) 2MASS measurements are available.   

\section{Spectroscopic observations}
\label{obs}
The stellar sample of G to M above was observed spectroscopically during several 
observing runs.
The journal of observations is given in Table \ref{journal}. Low-resolution spectra were 
obtained with CAFOS, high-resolution spectra were observed with FOCES, both attached to 
the 2.2\,m telescope at Calar Alto observatory (CA), Spain. 
Further high- and medium-resolution observations were obtained at the
Observatoire de Haute Provence (OHP), France, with the spectrographs AURELIE and CARELEC 
at the 1.52\,m and 1.93\,m telescopes, respectively. A few supplementary high- and 
low-resolution observations
were obtained at European Southern Observatory, La Silla, Chile (ESO), with CASPEC at 
the ESO 3.6\,m telescope and DFOSC at the Danish 1.54\,m telescope, respectively. 
A further observing run of 5 nights at Calar Alto observatory in February 2001 was  
lost due to bad weather conditions.

The spectra were reduced with the standard routines of the ESO-MIDAS software 
package. The low- and medium-resolution spectra and the
high-resolution spectra observed with  AURELIE were reduced with the 
{\em Longslit} package. 
For the FOCES and CASPEC data the routines of the {\em Echelle} package
were applied.

Spectra could be secured for the counterpart of 172  
out of 199 RASS sources 
with spectral types between G and M.  
High  resolution observations  were obtained for 118 of the 141 G and K stars 
of the selected sample (originally 143 G-K stars minus A031 and E020). 
Lithium equivalent widths and radial velocities for
six of the  stars not observed by us with high resolution 
were adopted from high-resolution spectroscopic studies by 
Wichmann et al. (\cite{Wichmannetal01}) (5 stars: A154, B049, B194, C062, C197)
and Neuh\"auser et al. (\cite{Neuhaeuseretal95}) (1 star: A058).
Ten G-K stars fainter than 12th magnitude were observed only with low resolution.
Thus for 134 of the 141 G-K stars spectroscopic follow-up observations exist.
For the remaining 7 stars no observations could be obtained.
Further high resolution data were found 
for the secondary counterpart of A098 in Favata et al. 
(\cite{Favataetal97}). 
With a few exceptions M stars were observed with low resolution only. Due to bad weather 
conditions during the OHP observing campaign the M stars in area V could not be observed.
In total 38 M stars were observed with low resolution and 7 with high resolution. For 13
M stars no observations could be obtained.

In the following we give more technical details of the spectroscopic observations.

\subsection{Low-resolution spectroscopy}
For the low-resolution  observations the focal reducer 
camera CAFOS attached to the 2.2\,m telescope at Calar Alto observatory, Spain, was used during
three observing runs. In 1996 and 1997 the instrument was
equipped with a LORAL-80 2048$\times$2048 pixel CCD chip with a pixel size of 
15$\mu$m. In 1998 a SITe1d 2048$\times$2048 pixel CCD chip with 24$\mu$m pixel size was 
used. Spectra in the wavelength range 4800--7450\,\AA\ were obtained (grism green-100)  
with a linear dispersion of 
1.3\,\AA\,px$^{-1}$ and 2.1\,\AA\,px$^{-1}$ 
with the LORAL and the SITe1d CCD chip, respectively.  With the LORAL chip the 
measured spectral resolution achieved with a 0.7\arcsec\ slit was 3.2\,\AA\ ($FWHM$). 
The SITe1d chip and a 1\arcsec\ slit yielded a spectral resolution of 4.2\AA .  
Several stars were additionally observed in the blue wavelength region between 
3850\,\AA\ and 5400\,\AA\  
with the grism b-100 and a 1\arcsec\ slit yielding similar spectral resolution as in the
red wavelength range. 
Wavelength calibration was obtained using He and HgRb lamps. For flat-field 
correction spectra of the dome illuminated with a halogen lamp were recorded. 

A few stars were observed in May 1998 with the focal reducer camera DFOSC attached to the 
Danish 1.54\,m telescope at ESO, La Silla. The spectra  were obtained with grism No. 7 
and a slit width of 1\arcsec . The wavelength range covered by the spectra was 3840--6845\,\AA .
As detector the LORAL/LESSER CCD\# C1W7 with a pixel size of 15\,$\mu$m was used. 
The resulting spectral resolving power was 1300.

\subsection{Medium-resolution  spectroscopy}
In May 1998 medium-resolution spectra were obtained with the spectrograph CARELEC 
(Lema\^{i}tre et al. \cite{Lemaitreetal90}) attached to the
Cassegrain focus of the 1.93\,m telescope at OHP. For the observations in the 
wavelength range
from 6420\,\AA\  to 6875\,\AA\ grating No. 2 with 1200\,lines\,mm$^{-1}$ was used in 1st order 
with a TEK CCD chip (pixel size 27$\mu$m). The linear dispersion was 33\AA\,mm$^{-1}$. 
The spectral resolution achieved was about 4600.  

\subsection{High-resolution  spectroscopy}
The largest part of the high-resolution observations were obtained during four observing
campaigns with the echelle spectrograph FOCES (cf. Pfeiffer et al. \cite{Pfeifferetal98}) 
at the 2.2\,m telescope of Calar Alto Observatory. 
The spectrograph was coupled to the telescope with the
red fibre. The detector was a 1024$\times$1024 pixel Tektronix 
CCD chip with 24\,$\mu$m pixel size. With a diaphragm diameter of 
200\,$\mu$m and an entrance slit width of 180\,$\mu$m a spectral resolution of
34\,000 was achieved. Wavelength calibration was obtained with a ThAr lamp. The nominal
spectral coverage is from 3880\AA\ to 6850\AA . However, due to the wavelength
dependence of the transmission curve of the red fiber and the continuum energy distribution of
the stars the useful spectral range 
of the spectra is typically from $\sim5000$\AA\ to 6850\AA . At shorter wavelength the
S/N ratio decreases.

In October 1998 high-resolution spectra were obtained with the spectrograph AURELIE at the 
1.52\,m telescope of the OHP. A description of the spectrograph can be found in Gillet et 
al. (\cite{Gilletetal94}). The spectra were observed with grating No. 2
with 1200\,lines\,mm$^{-1}$ giving a reciprocal linear dispersion of 8\,\AA\,mm$^{-1}$.
The detector was a double-barrette Thomson TH7832 (2048 pixel with 
13\,$\mu$m pixel size). The spectra cover the wavelength interval from
6540\,\AA\ to 6740\,\AA . The resolution of the spectra is 20\,000. Wavelength
calibration was obtained with Neon and Argon lamps.

High-resolution spectra of 3 objects were obtained  with the Cassegrain Echelle
Spectrograph (CASPEC) at the ESO 3.6\,m telescope on La Silla in 
February 1998. Wavelength calibration was obtained with a ThAr lamp. 
The  CASPEC spectra cover the spectral range from 5350 to 7720\,\AA\  
with a nominal resolving power of  22,000 (Sterzik et al. \cite{Sterziketal99}).

During each high-resolution observing campaign radial and rotational velocity 
standard stars were observed in addition to the science targets.

\section{Observational results}
\label{obsresult}
\subsection{Spectral classification}
\label{specclass}
In Paper III spectral types were given based largely on low-resolution
classification spectra obtained with LFOSC (cf. Paper II). For a  smaller number of 
stars spectral types were adopted 
from the literature. Our high-resolution spectra not only allowed us to refine the 
classification but, even more importantly, enabled us to  derive luminosity 
classes and hence spectroscopic parallaxes. 

During the observing runs a small set of spectroscopic standard stars, mainly of
luminosity class V, had been observed together with the science targets. The coverage of the spectral type -
luminosity class plane, however, was insufficient for a detailed two-dimensional classification. We
therefore extended the spectroscopic data base for the standard stars by making use of
the spectra available in the stellar 
library\footnote{URL: http://www.obs-hp.fr/www/archive/archive.html}  of Prugniel \& Soubiran  
(\cite{PrugnielSoubiran01}) which is part of the HYPERCAT\footnote{URL:
http://www-obs.univ-lyon1.fr/hypercat/} data base.
We used the data set  with a spectral resolution of 10\,000. In
order to match this resolution  our FOCES, AURELIE, and CASPEC spectra were smoothed
accordingly with an appropriate Gaussian filter. In this way the
signal-to-noise ratio  improved while the necessary spectral resolution for the classification 
was preserved. Spectral types and luminosity classes (LCs) of MK standard stars contained in 
the stellar library were adopted from 
Yamashita et al. (\cite{Yamashitaetal76}),  Keenan \& McNeil (\cite{KeenanMcNeil89}), 
Garcia (\cite{Garcia89}), Keenan \& Barnbaum (\cite{KeenanBarnbaum99}), and 
Gray et al. (\cite{Grayetal01}). In a few cases we adopted the spectral classification
given in  Prugniel \& Soubiran  (\cite{PrugnielSoubiran01}). The grid of spectroscopic 
standard stars is listed in Table \ref{specstd}.

In a pilot study for the work presented here Ziegler (\cite{Ziegler93}) studied the spectral 
types of F, G and K-type stars from the RASS using spectra observed in the red spectral 
region ($\lambda\lambda$6200 - 6750\,\AA ). He found various line ratios useful for 
classification purposes. For the F- and G-type stars the 
ratios \FeI $\lambda$6394/\SiII $\lambda$ 6346, \FeII  $\lambda$6456/\CaI  $\lambda$6450 and 
\FeII $\lambda$6456/ \FeI $\lambda$6394 were found
to be good indicators for the spectral type. In K stars the ratios TiO 
$\lambda$6240 / \VI $\lambda$6296
and \FeI$\lambda$ 6250/\CaI $\lambda$6450 were useful classification criteria. 

\begin{figure}[tb]
\resizebox{\hsize}{!}{\includegraphics[angle=-90,bbllx=50pt,bblly=50pt,bburx=550pt,bbury=715pt,clip=]{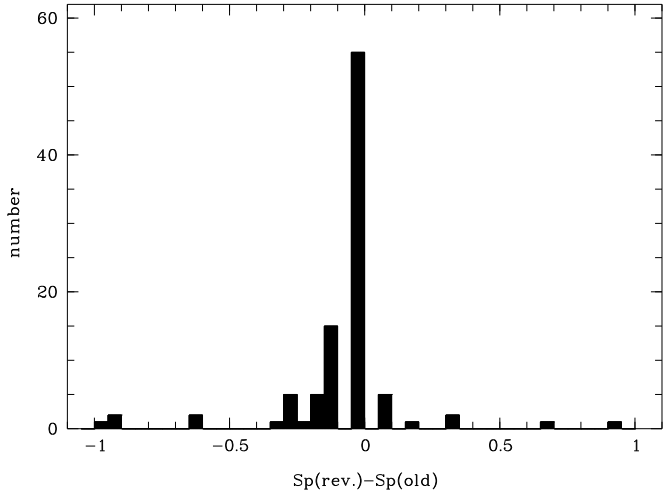}}
\caption[]{Comparison of the spectral types derived from the classification spectra
used in Paper III (Sp(old)) and from the new high resolution spectra (Sp(rev.)). 
The abscissa is the difference (in spectral classes) between the revised and the original 
spectral types.
}
\label{speccomp}
\end{figure}

We used these ratios for the refinement of the spectral types given in Paper III.
Figure \ref{speccomp} shows the histogram of the differences between the revised  and 
original spectral types. The narrow peak shows that with few
exceptions the overall agreement is good. We found a small mean difference 
of -0.5 subclass between the high- and low-resolution spectral types with
a standard deviation of 2.2 subclasses. The original and the revised statistics of 
spectral types are listed in  Table \ref{revstat}. In nine cases the difference of the spectral
types was larger than $\pm3$ subclasses. The largest differences were found for B174 and E256
($-6$ subclasses), B185 (7 subclasses), D018 (9 subclasses), and E022 and E067 ($-9$ subclasses).
The LFOSC spectrum of E256 was actually classified as K4, 
but erroneously entered in Paper III as M0. For D018 which is a very bright star the original 
LFOSC spectrum classified as G2V could suffer from saturation. In SIMBAD  this star is listed as K0III
(Schild \cite{Schild73}). 
The classification based on the FOCES spectrum is K1III, which is in good agreement 
with the literature. We adopt this spectral class in the following. For the remaining stars with large
deviations no LFOSC classification spectra were obtained. The spectral classes were adopted 
from SIMBAD. In the following we use the improved FOCES classifications.

\begin{table} 
\centering
\caption[]{Revised and original  statistics  of the distribution of 
spectral type among the RASS sources with stellar counterparts of 
spectral types F to M.
} 
\begin{tabular}{lll} 
\noalign{\smallskip}     
\hline 
\hline 
\noalign{\smallskip}    
spec. type   & this work &  Paper IV\\
\hline 
\noalign{\smallskip}    
F                &  55  &  53 \\ 
G                &  56  &  54 \\ 
K                &  86  &  89 \\ 
M                &  56  &  58 \\ 
\hline 
\noalign{\smallskip}    
total        &  253   &  254 \\
\label{revstat}
\end{tabular}
\end{table}

Following Gahm \& Hultqvist (\cite{GahmHultqvist72}) and Ziegler (\cite{Ziegler93})  
luminosity classes (LC) were obtained using the strength of the lines of 
\BaII$\lambda\lambda$5854\AA , 6497\AA , \ScII$\lambda$6605\AA ,  and 
\LaII$\lambda$6390\AA . We added the \YII$\lambda$6614\AA\ line which also shows a clear
luminosity dependence. 
The ratio of \ScII$\lambda$6605\AA\ and \YII$\lambda$6614\AA\ is a good 
luminosity indicator for spectral types earlier than about K5-7. For spectral types 
later than K0 the
strength of \LaII\ was additionally useful to discriminate luminosity classes III and 
higher from LC V and IV. For G stars LC III and higher could also be discriminated 
from LC IV by the use of this line.
Comparing in this way the line strengths and ratios in the MK standards 
with the sample stars LCs could be assigned to most stars. For a few stars the stellar 
absorption lines were strongly broadened by rapid rotation (see below). In these cases it
was  not possible to determine the luminosity class due to the 
limited S/N of the spectra and to line blending. The limit was reached around 
$v\sin i \ga 30$\,\kms . For the rapid rotators we adopted LC V. As discussed in
Sect. \ref{distance} we used the luminosity classes to derive spectroscopic 
parallaxes.

\begin{table} 
\centering
\caption[]{Spectroscopic MK standard stars. The spectral types are listed in column 
``sp. type''. References for the spectral types are given in column ``ref.'': 1 = Yamashita et al.
(\cite{Yamashitaetal76}),
2 = Gray et al. (\cite{Grayetal01}), 3 = Keenan \& Barnbaum (\cite{KeenanBarnbaum99}),
4 = Garcia (\cite{Garcia89}), 5 =  Keenan \& McNeil (\cite{KeenanMcNeil89}), 6 = Prugniel \& Soubiran  
(\cite{PrugnielSoubiran01}).
} 
\begin{tabular}{llllll} 
\noalign{\smallskip}     
\hline 
\hline 
\noalign{\smallskip}    
star       &  sp. type & ref. & star       &  sp. type & ref.\\
\hline 
\noalign{\smallskip}    
\object{HD\,222368} & F7V    &   4 & \object{HD\,188119} & G7III  &   4  \\
\object{HD\,016765} & F7IV   &   6 & \object{HD\,010700} & G8V    &   4  \\
\object{HD\,216385} & F7IV   &   6 & \object{HD\,188512} & G8IV   &   4  \\
\object{HD\,181214} & F8III  &   6 & \object{HD\,027348} & G8III  &   3  \\
\object{HD\,004614} & G0V    &   4 & \object{HD\,175306} & G9III  &   4  \\
\object{HD\,013974} & G0V    &   4 & \object{HD\,145675} & K0V    &   5  \\
\object{HD\,019373} & G0V    &   4 & \object{HD\,185144} & K0V    &   4  \\
\object{HD\,114710} & G0V    &   1 & \object{HD\,198149} & K0IV   &   5  \\
\object{HD\,150680} & G0IV   &   4 & \object{HD\,048433} & K0III  &   3  \\
\object{HD\,039833} & G0III  &   6 & \object{HD\,010476} & K1V    &   5  \\
\object{HD\,204867} & G0Ib   &   1 & \object{HD\,222404} & K1IV   &   5  \\
\object{HD\,204613} & G1III  &   4 & \object{HD\,096833} & K1III  &   5  \\
\object{HD\,185758} & G1II   &   4 & \object{HD\,022049} & K2V    &   4  \\
\object{HD\,186408} & G2V    &   4 & \object{HD\,137759} & K2III  &   4  \\
\object{HD\,126868} & G2IV   &   2 & \object{HD\,020468} & K2II   &   4  \\
\object{HD\,209750} & G2Ib   &   1 & \object{HD\,219134} & K3V    &   4  \\
\object{HD\,117176} & G4V    &   5 & \object{HD\,003712} & K3III  &   3  \\
\object{HD\,127243} & G4IV   &   5 & \object{HD\,201091} & K5V    &   4  \\
\object{HD\,186427} & G5V    &   1 & \object{HD\,118096} & K5IV   &   6  \\
\object{HD\,161797} & G5IV   &   4 & \object{HD\,029139} & K5III  &   3  \\
\object{HD\,027022} & G5IIb  &   5 & \object{HD\,088230} & K6V    &   4  \\
\object{HD\,206859} & G5Ib   &   1 & \object{HD\,201092} & K7V    &   4  \\
\object{HD\,003546} & G6III  &   5 & \object{HD\,079210} & M0V    &   6  \\
\object{HD\,182572} & G7IV   &   4 & \object{HD\,046784} & M0III  &   6  \\
\hline 
\noalign{\smallskip}    
\end{tabular}
\label{specstd}
\end{table}

\subsection{Radial and rotational velocities}
During each observing run for high-resolution spectroscopy a set of radial and rotational velocity 
standards had been observed together 
with the RASS counterparts. Heliocentric radial velocities were measured  by means of a
cross-correlation method (Simkin \cite{Simkin74}). The continuum was subtracted from the normalized spectra 
which were then rebinned on a logarithmic wavelength scale. The shift  relative to the radial 
velocity standards was measured and transformed into the radial velocity of the target by taking 
into account the radial velocity of the standard  stars. The  individual radial velocities obtained for 
each standard star were averaged to give the final result.
The standard deviation gives a measure for the error. With few exceptions 
(spectra with low S/N and/or high
rotational velocity) the errors were in the range 1-4\,\kms\ with a 
typical error of about 2-3\,\kms .
Heliocentric radial velocities (and errors) are listed in Table \ref{resulttab2}. 

\begin{figure}[tb]
\resizebox{\hsize}{!}{\includegraphics[angle=-90,bbllx=45pt,bblly=45pt,bburx=540pt,bbury=770pt,clip=]{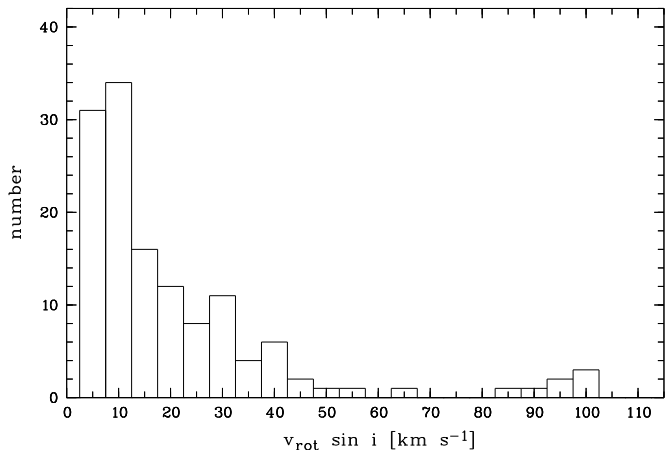}}
\caption[]{Distribution of rotational velocities of the G and K stars.
}
\label{vrothist}
\end{figure}

The width of the cross-correlation function is a measure for the rotational 
velocity $v\,\sin i$. We therefore calculated the cross-correlation function as 
before but for rotational velocity standards.
Standard stars with low $v\,\sin i$ and spectral type as
close as possible to that of the objects were used for the cross-correlation analysis 
as well as to calibrate the FWHM vs.  $v\,\sin i$ relation. From the FWHM of the 
cross-correlation function $v\,\sin i$ was then determined following the method described 
in Covino et al. (\cite{Covinoetal97}). 
Observations of rotational standard stars yielded a detection limit of $v\,\sin i$ of 
about 5\,\kms . From the statistics of the differences between
measured rotational velocities of rotational standard stars and $v\,\sin i$ from the 
literature an uncertainty of $v\,\sin i$ of 3\,\kms\ could be estimated.
For rotational velocities above $\sim40$\,\kms\ the shape of the peak of the 
correlation function deviates increasingly from a Gaussian leading to larger 
errors of 5-10\,\kms . 
Figure \ref{vrothist} shows the histogram of the rotational velocities which 
are listed in Table \ref{basic}. 

\subsection{Lithium equivalent widths}
\label{lisurvey}
Equivalent widths (EWs) of the lithium absorption line \LiI$\lambda6708$, $W$(\LiI ) 
were determined from the low-, medium-  and high-resolution spectra.
The measurement of the EW  in the low- and medium re\-so\-lution spectra was performed as described in 
detail in Paper VI. Essentially, the method takes the line blending with neighboring \FeI\ lines into account 
by fitting Gaussian profiles at the wavelengths of the \FeI\ lines at 6703, 6705, and 
6710\,\AA\ simultaneously with the lithium line at 6708\AA . In Paper VI the error of 
$W$(\LiI ) determined from the CAFOS spectra was estimated to be about 60\,m\AA . For the DFOSC
spectra the uncertainty is similar. The fitting procedure was also applied to the 
medium-resolution CARELEC spectra. The uncertainty of the EW for these spectra is about 
40\,m\AA . 

In the high-resolution spectra the equivalent widths were measured directly by integrating the
flux in the normalized spectra. The contribution of the neutral iron line 
\FeI$\lambda6707.441$\AA\ was corrected according to the procedure described by Soderblom 
et al. (\cite{Soderblometal93b}). For stars with rotational velocities 
larger than
$\sim30$\,\kms\ the contribution of the \FeI\ lines near \LiI$\lambda6708$ was
corrected in the following way. 
From the stellar library of Prugniel \& Soubiran  a spectroscopic standard star
with a spectral type as close as possible to the target was selected. It was folded 
with the appropriate rotational velocity to match the broadened lines of the target
spectrum. Then the EW of the \FeI\ absorption features  was measured
in the same wavelength interval as used to determine the \LiI\ EW in the
target spectrum. Finally the corrected lithium EW was obtained by subtracting 
the contribution of the \FeI\ lines from the measured  
lithium EW of the target spectrum. 
Errors of the high-resolution EWs are typically 5-15\,m\AA , 
depending on the
signal-to-noise ratio and on the rotational velocity.
The EWs are listed in Table \ref{resulttab3}.

\begin{figure}[tb]
\resizebox{\hsize}{!}{\includegraphics[angle=-90,bbllx=53pt,bblly=25pt,bburx=552pt,bbury=542pt,clip=]{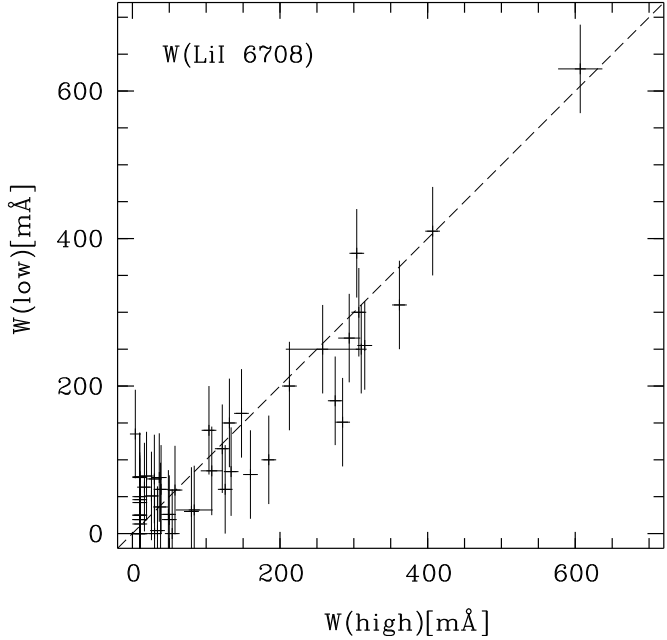}}
\caption[]{Comparison of the equivalent widths of \LiI\ determined from the low- and the
high-resolution spectra. The dashed line denotes a ratio of 1 of the two measurements.
}
\label{wlihilow}
\end{figure}

In Fig. \ref{wlihilow}  the EWs obtained from the low- and
the high-resolution spectra are compared. In the low-resolution spectra the 
EWs $W$(\LiI ) are obviously slightly underestimated by about 40\,m\AA . However, 
the overall agreement is good 
and the differences are only of the order the uncertainty of the low-resolution measurements. 
This demonstrates that the fitting method applied to the low-resolution spectra works 
remarkably well. In particular, $W$(\LiI ) is not overestimated as it would be the case 
if the EWs would be determined directly by flux integration without taking the contribution 
of the \FeI\ lines into account. 

\subsection{Binaries}
Spectroscopic binaries were detected by means of the shape of the cross-correlation function 
obtained for the radial velocity determination. Among the 125 G-K stars with 
high-resolution spectroscopy either obtained by us or taken from the literature 
32 binary systems and 1 triple system were found. The triple system is B160. The fraction of multiple
systems in our sample of G-K type stars with high-resolution observations thus is 26\% with a
lower limit of 23\% for the full sample of G-K stars. 

In a few binaries lithium lines could be identified in one or both components. In order to
disentangle the lines of the individual components and to identify a possible \LiI\ line
spectra from the Prugniel \& Soubiran 
sample with the appropriate spectral types were folded with the rotational profile for the
measured $v\,sin i$ and shifted with respect to the measured radial velocities. Then
the spectra were superimposed by using appropriate values for the relative flux contributions.
Finally the resulting artificial binary spectrum was compared with the observed spectrum. 
Correction factors for the measured lithium equivalent widths were estimated from the artifical
spectrum. In most cases the spectra suggest a flux ratio of 1 to 2 for the individual 
components at 6708\,\AA .
Exceptions are e.g. A001 and A071.  In A001 the primary component is a fast rotator ($v\,sin i
\approx 100$\,\kms) whose broad lines dominate the spectrum.
Of the sec\-on\-dary component only the strongest lines of a mid to late type K star are detectable.
For this binary system we adopted a flux ratio of 5:1 for the continuum contributions of the primary and
secondary component at 6708\,\AA . In A071 both components are fast rotators with very broad lines.
In this case it was not possible to determine a lithium EW for each  component. 
The total EW was therefore assigned in equal shares to the individual components and 
the lithium equivalent widths were corrected by assuming equal flux
contributions. 
The triple system B160 is even more complicated. It consists of 3 early to mid G-type stars with
spectral types between $\sim$G2 and $\sim$G5. 
Two of the three components exhibit a lithium absorption line.  

It is clear that the
equivalent widths of the binaries and the triple system are less reliable than those of the
single stars due to the uncertainty of the continuum correction. 
In Table \ref{resulttab3} 
the lithium EW of the strongest component is given.

\section{Data analysis and discussion}
\label{analysis}
In the following we will first discuss the basic parameters of the coronal 
sample and then investigate the age distribution using lithium abundances, 
and the kinematics as derived from radial velocities and proper motions.

\subsection{Basic properties}
\subsubsection{Distances}
\label{distance}
The distance is clearly one of the most important parameters. 
For 58 of the 252 F-M type counterparts  a Hipparcos parallax with 
$\pi_{\rm H}/\sigma_{\rm H} \ge 3$ exists.  
The 58 stars with Hipparcos parallax comprise 28 F stars, 17 G stars, 
9 K stars and 4 M stars. Further trigonometric parallaxes of 7 M stars were found 
in  Gliese \& Jahreiss  (\cite{GlieseJahreiss91}). 

\begin{figure}[tb]
\resizebox{\hsize}{!}{\includegraphics[angle=0,bbllx=60pt,bblly=50pt,bburx=545pt,bbury=600pt,clip=]{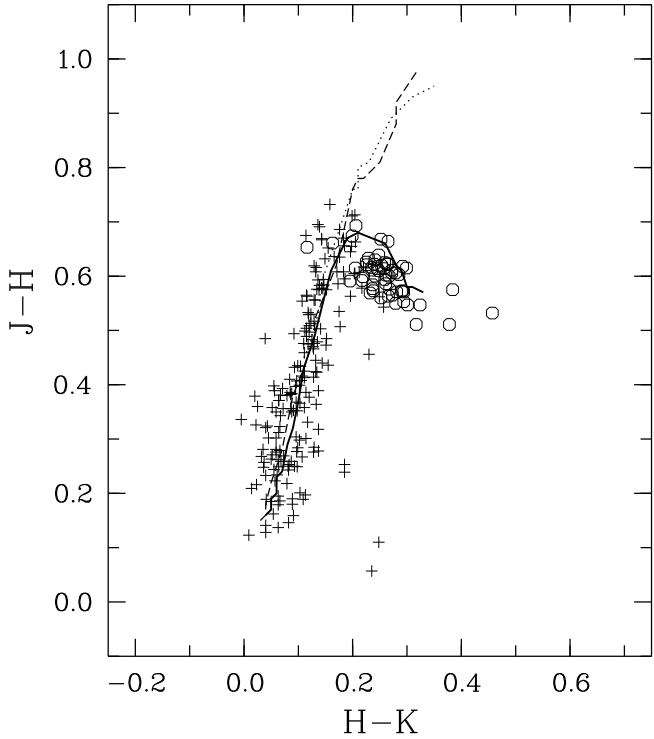}}
\caption[]{Two-colour diagram for the infrared magnitudes from 2MASS. Circles denote 
M stars, crosses stars with spectral types F  to K. The solid, dotted, and dashed lines denote the loci of main sequence 
stars, giants, and supergiants, respectively. 
}
\label{jhhk}
\end{figure}

For 74 stars a spectroscopic parallax could be derived from the high-resolution spectra 
by adopting the absolute $V$ magnitudes, as appropriate for the spectroscopically 
determined luminosity class, from Schmidt-Kaler (\cite{SchmidtKaler82}).  
For the bulk of M stars we used infrared $JHK$ measurements from the 2MASS catalogue 
to derive a photometric distance. The two-colour diagram of $J-H$ and $H-K$ is displayed 
in Fig. \ref{jhhk}. It shows that the M stars  
are distributed around the locus of main-sequence stars (solid line in 
Fig.\ref{jhhk}). For the further analysis distances of M stars 
were therefore estimated by adopting $M_V$ for LC V from Schmidt-Kaler 
(except for the 11 stars with trigonometric parallaxes). 
This adds 43 more RASS sources with a distance estimate. 
Thus total distances are available for 100 G-K and 54 M stars.
For the remaining
stars without a distance measurement we derived a lower limit for the 
distance by assuming that they are main-sequence objects  with LC V.

An estimate of the error of the spectroscopic and photometric distances, $\sigma_d$, may be obtained
from the following considerations. The error is due to the uncertainties of the absolute
visual magnitude, $M_V$, and  of $V$. For the latter we conservatively adopted the error of the 
photographic GSC magnitudes $\sigma_V = 0.3^{\rm m}$ for all stars. The dominating source of
uncertainty is the error of $M_V$. For G-K stars of LC V and IV  and 
correspondingly for LC III and II we used half of the
difference of $M_V$ of these luminosity classes as estimate for $\sigma_{M_V}$. 
This leads to an estimate for $\sigma_d/d$ of 30-50\%. In the case of M stars the main
source of error of $M_V$ is due to the uncertainty of the spectral class. This also leads
in total to $\sigma_d/d \sim$50\% if an uncertainty of 1-2 spectral subclasses is 
assumed. We finally adopted 50\% as relative error for spectroscopic and 
photometric distances.

For the derivation of the distances interstellar extinction was 
not taken into account. Given the high galactic latitude of our sample it is 
actually expected to be small. With the relation $N_H = 5.9\,10^{21} 
\times E(B-V)$ given by Spitzer (\cite{Spitzer78}) with the column
density of neutral hydrogen, $N_H$, and colour excess $E(B-V)$ upper limits of the
extinction can be estimated.  We expect extinction
values, $A_V$, of less than 0.2-0.3 in all study areas except area I. This region 
could have a higher extinction of up to 0.6 magnitudes for the most distant stars. 
For these estimates the $N_H$ values given in Paper II were used. 

\begin{figure}[tb]
\resizebox{\hsize}{!}{\includegraphics[angle=-90,bbllx=80pt,bblly=30pt,bburx=565pt,bbury=515pt,clip=]{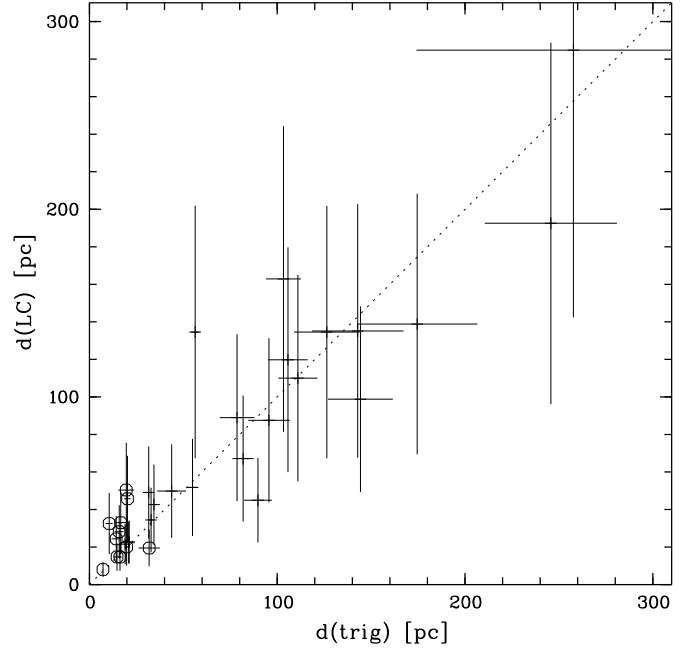}}
\caption[]{Comparison of the distances $d({\rm LC})$ derived from spectroscopic or 
photometric  parallaxes and from trigonometric parallaxes, $d({\rm trig})$. 
The dashed line denotes equal distance values. M stars with distance estimates
from the 2MASS IR photometry are plotted as circles.
}
\label{distcomp}
\end{figure}

For 20 stars in our sample both spectroscopic and Hipparcos parallaxes, 
$\pi_{\rm H}$, exist. They are compared in Fig. \ref{distcomp}. 
The agreement of the two distance measurements 
for this subsample is good. The mean ratio of both parallaxes is $1.06\pm0.35$. 
For the further analysis we adopted the spectroscopic parallaxes if no Hipparcos 
parallax with  $\pi_{\rm H}/\sigma_{\pi} > 3$ or other trigonometric
parallax was available. The adopted distances are listed in 
Table \ref{basic}. 

\begin{figure}[tb]
\resizebox{\hsize}{!}{\includegraphics[angle=-90,bbllx=40pt,bblly=30pt,bburx=570pt,bbury=770pt,clip=]{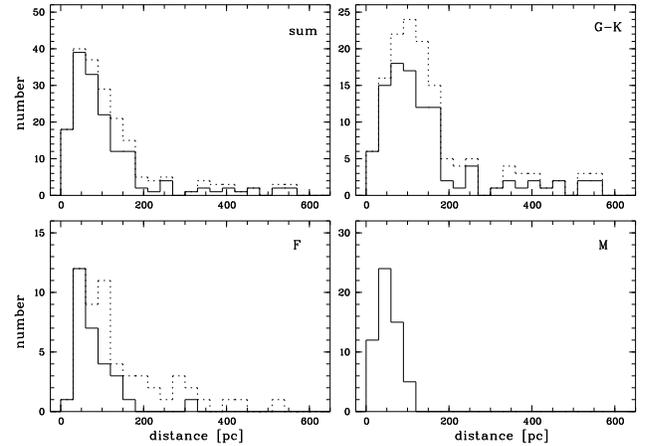}}
\caption[]{Histogram of the distance distribution. The solid lines represent the
distribution of trigonometric, spectroscopic, and photometric parallaxes. 
The dashed lines include distance estimates derived from assuming  absolute 
visual magnitude of main-sequence stars for the remaining stars without other distance
estimate.  
}
\label{disthist}
\end{figure}

Figure \ref{disthist} shows the number distribution of the distances for the 
184 F-, G-, and M stars. Also shown is the distribution  including the 
stars with minimum distances estimated by adopting LC V.
The number distribution of the total sample has a maximum around 50\,pc 
with a tail extending up to several 100\,pc. Most stars are nearer than 200\,pc, 
33 stars have distances above 300\,pc (including 16 stars with 
minimum distances), and in 4 cases (not shown in Fig. \ref{disthist}) we derived 
a distance above 1\,kpc (including 3 stars with minimum distances). The identifications 
of the very distant RASS counterparts may be questionable. 

For the stars with trigonometric parallaxes  the
absolute magnitude, $M_V$, was calculated from the distance and visual magnitude 
given in Table \ref{basic}. A luminosity class was then assigned according to 
Schmidt-Kaler (\cite{SchmidtKaler82}). Likewise, bolometric corrections were taken 
from the same reference to determine the bolometric magnitudes for all stars with
known distances.

\begin{figure}[tb]
\resizebox{\hsize}{!}{\includegraphics[angle=-90,bbllx=80pt,bblly=60pt,bburx=540pt,bbury=790pt,clip=]{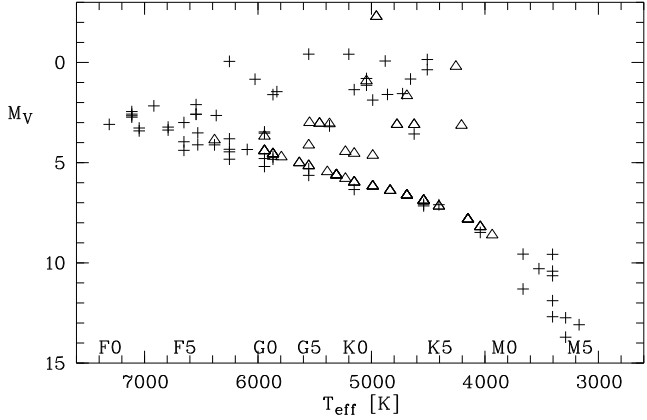}}
\caption[]{H-R-diagram for single stars with either a trigonometric parallax from Hipparcos
or other sources ($+$ sign) or with a 
spectroscopic parallax (triangles). 
}
\label{MVspec}
\end{figure}

As expected the majority of stars with a luminosity class determination, $\sim$90\%,
have luminosity class V or IV. A small number of 17 stars was classified as giants 
(LC III-IV, III, and II), 12 of these based on Hipparcos parallaxes.
In Fig. \ref{MVspec}  the H-R diagram is shown for 
all stars with a spectroscopic or trigonometric parallax. 
M stars are shown only  if a trigonometric parallax 
was available. 

\subsubsection{X-ray properties}
\begin{figure}[tb]
\resizebox{\hsize}{!}{\includegraphics[angle=-90,bbllx=80pt,bblly=60pt,bburx=545pt,bbury=790pt,clip=]{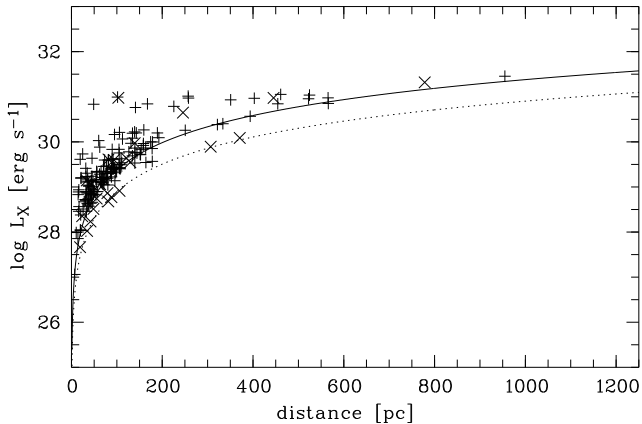}}
\caption[]{X-ray luminosity for single stars 
vs. distance. $+$ signs mark stars in study areas I, II, III, IV, and VI, 
$\times$ signs represent stars in area V. 
The solid and dotted lines mark the flux limits for the two groups of study areas.
}
\label{distlx}
\end{figure}

In Paper II we discussed the  X-ray flux limits in the ROSAT 0.1--2.4\,keV energy band 
for the various classes of X-ray emitters in our sample. For coronal emitters it is 
2\,10$^{-13}$\,erg\,cm$^{-2}$\,s$^{-1}$. An exception is study area V which  
due to the deeper RASS exposure near the north ecliptic pole has a lower flux limit 
of 0.6\,10$^{-13}$\,erg\,cm$^{-2}$\,s$^{-1}$. 
X-ray luminosities, $L_{\rm X}$, were derived from the fluxes
given in Paper III and the distances derived here. In Fig. \ref{distlx}  
$L_{\rm X}$ is plotted vs. the distance. 
Also shown are the two flux limits.
As expected for a flux-limited sample this
plot shows a correlation between distance and luminosity because at increasingly
larger distances only the  more luminous objects are detected. 

\begin{figure}[tb]
\resizebox{\hsize}{!}{\includegraphics[angle=-90,bbllx=80pt,bblly=60pt,bburx=545pt,bbury=790pt,clip=]{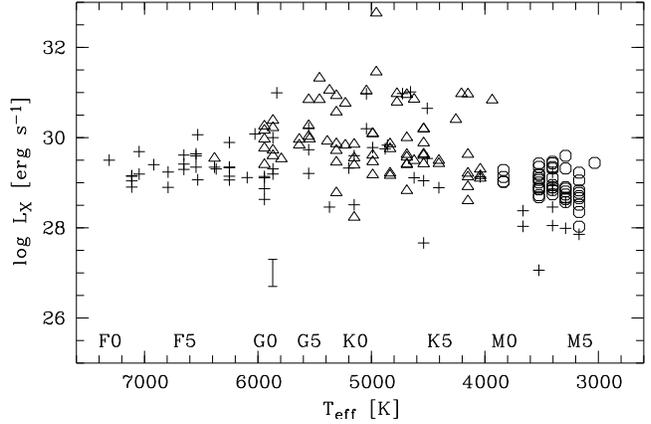}}
\caption[]{X-ray luminosity for single stars as a function of effective 
temperature, $\teff$. Different symbols  identify stars with trigonometric 
($+$ sign), spectroscopic (triangles), or IR photometric parallaxes (circles). 
The variability range of solar X-ray emission in the ROSAT-PSPC pass band
is marked by the vertical bar.
}
\label{loglumx}
\end{figure}

\begin{figure}[tb]
\resizebox{\hsize}{!}{\includegraphics[angle=-90,bbllx=80pt,bblly=60pt,bburx=540pt,bbury=790pt,clip=]{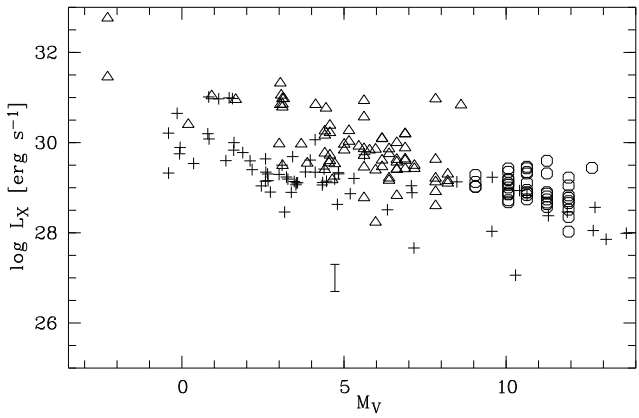}}
\caption[]{X-ray luminosity for all single stars with trigonometric ($+$ sign), spectroscopic (triangles), or IR photometric parallaxes 
(circles)  as a function of absolute visual magnitude, $M_V$. The
variability range of solar X-ray emission is marked by the vertical bar.
}
\label{MVLx}
\end{figure}

In Fig. \ref{loglumx} 
$L_{\rm X}$ is plotted versus the effective temperature and Fig. \ref{MVLx} 
shows $L_{\rm X}$ as function of the absolute visual magnitude, $M_V$. A weak trend 
of $L_{\rm X}$ increasing with increasing $\teff$ is visible. 
The  $L_{\rm X}$-$M_V$  diagram shows a clear correlation with $L_{\rm X}$ 
decreasing for decreasing optical luminosity. This reflects the fact that 
$L_{\rm X}$ depends on the emitting surface. 
The width of the $L_{\rm X}$-$M_V$ distribution at a given $M_V$ tells 
that the  X-ray surface flux density of the stars in our sample spans a range of 
a factor of $\sim1000$. Around $M_V = 5$ the lower limit of the X-ray luminosities of 
the sample stars is about a factor of 10 above the solar soft X-ray variability range 
($5\,10^{26}-2\,10^{27}$\,erg\,s$^{-1}$,  Schmitt \cite{Schmitt97}). The upper limit of
$L_{\rm X}$ in our sample is about a factor of 10-30 higher than in the volume-limited
sample of Schmitt (\cite{Schmitt97}).

\begin{figure}[tb]
\resizebox{\hsize}{!}{\includegraphics[angle=-90,bbllx=80pt,bblly=60pt,bburx=550pt,bbury=790pt,clip=]{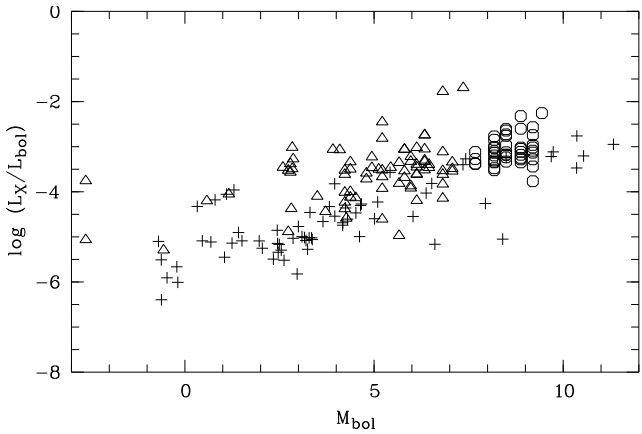}}
\caption[]{Ratio of X-ray and bolometric luminosity for all single stars with
trigonometric ($+$ sign), spectroscopic (triangles), or IR photometric parallaxes 
(circles) as a function of bolometric magnitude, $M_{\rm bol}$. 
}
\label{loglxlbol}
\end{figure}

The ratio of $L_{\rm X}$  and bolometric luminosity, $L_{\rm bol}$, is plotted in Fig. 
\ref{loglxlbol} as function of $M_{\rm bol}$. A clear correlation is visible with the
low luminosity stars with later spectral types having the highest ratio of 
$L_{\rm X}/L_{\rm bol}$.
This is in agreement with the results of Fleming et al. (\cite{Flemingetal95}) who 
studied the coronal X-ray activity of low-mass stars in a volume limited sample.
They found the highest ratios of 
$L_{\rm X}/L_{\rm bol}$ for dMe stars. 
As discussed in Paper IV, most M stars in our sample
are actually dMe stars, that is of the 58 M stars listed originally in Paper III 53 
exhibit \Ha\ emission lines. Note, however, that selection effects inherent 
in our flux-limited sample may also play a role.

\begin{figure}[tb]
\resizebox{\hsize}{!}{\includegraphics[angle=-90,bbllx=80pt,bblly=60pt,bburx=545pt,bbury=790pt,clip=]{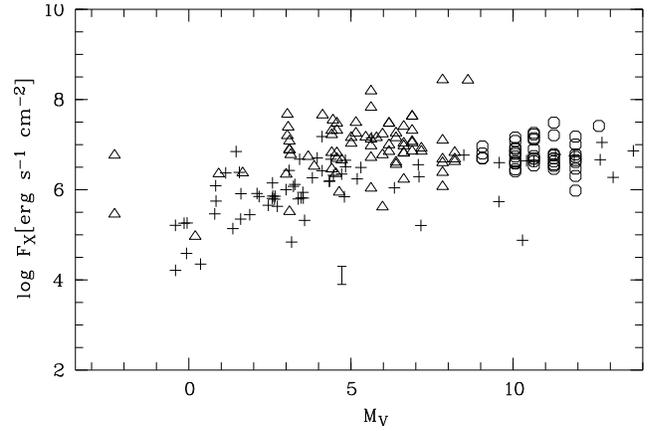}}
\caption[]{X-ray surface flux density for all single stars with trigonometric ($+$ sign), spectroscopic (triangles), 
or IR photometric parallaxes (circles)  as a function of absolute visual 
magnitude, $M_V$. The vertical bar marks
the typical flux level of  solar coronal holes in the ROSAT-PSPC pass band.
}
\label{FxMV}
\end{figure}

\begin{figure}[tb]
\resizebox{\hsize}{!}{\includegraphics[angle=-90,bbllx=80pt,bblly=60pt,bburx=545pt,bbury=790pt,clip=]{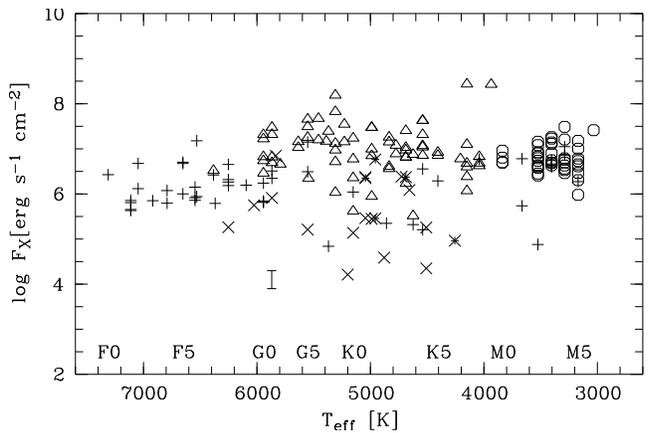}}
\caption[]{X-ray surface flux density for all single stars
as a function of effective temperature, $\teff$.  The meaning of the symbols is the
same as in Fig. \ref{loglumx}.
The vertical bar marks
the typical flux level of  solar coronal holes in the 
ROSAT-PSPC pass band.
}
\label{FxTeff}
\end{figure}

The X-ray surface flux density is displayed as a function of $M_V$  in Fig. \ref{FxMV} and
as a function of $\teff$ in Fig. \ref{FxTeff}. Our sample contains 
mainly stars with a high surface flux density which is on the average 1 to 2 orders of
magnitude above the solar flux level. This can be understood in view of the result 
discussed below in  Sect. \ref{agegroups} that our sample contains a large
fraction of young and hence very X-ray active stars.   
Old  solar-like stars are obviously not present in our sample. The maximum value 
of the surface flux density of our sample stars is around 
$10^8$\,erg\,s$^{-1}$\,cm$^{-2}$. This value is consistent with the result obtained by
Schmitt (\cite{Schmitt97}) who found a maximum around 
$10^7-10^8$\,erg\,s$^{-1}$\,cm$^{-2}$ in his volume-limited sample of solar-like stars.

\begin{figure}[tb]
\resizebox{\hsize}{!}{\includegraphics[angle=-90,bbllx=80pt,bblly=40pt,bburx=562pt,bbury=770pt,clip=]{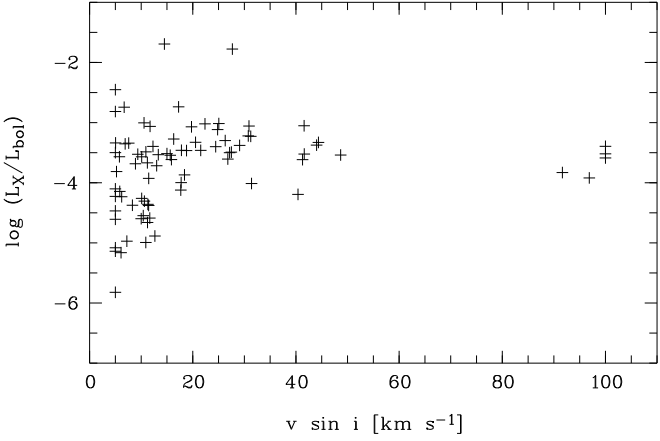}}
\caption[]{$\log L_X/L_{\rm bol}$ as a function of $v\,\sin i$. 
}
\label{lxvrot}
\end{figure}

Finally, in Fig. \ref{lxvrot} the ratio $\log L_X/L_{\rm bol}$ is displayed as 
function of
projected rotational velocity, $v\,\sin~i$. 
No clear correlation can be seen, except that small ratios of $\log L_X/L_{\rm bol}$ are only found
for small $v\,\sin~i$, whereas fast rotators exhibit high $\log L_X/L_{\rm bol}$ ratios.

\subsection{Lithium abundances and age distribution}

\subsubsection{Lithium abundances}
\begin{figure}[tb]
\resizebox{\hsize}{!}{\includegraphics[angle=-90,bbllx=82pt,bblly=56pt,bburx=547pt,bbury=773pt,clip=]{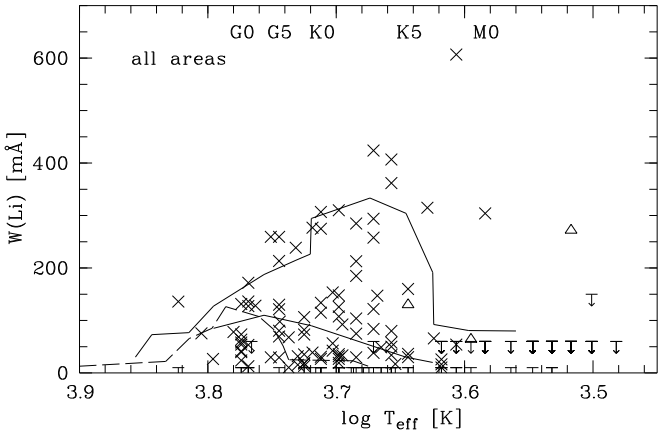}}
\caption[]{Equivalent widths of \LiI$\lambda6708$ as a function of $\teff$ for all stars in the six study
areas. Crosses and triangles are high and low resolution measurements, respectively. 
The solid lines represent the upper and lower envelope of the
lithium equivalent widths in the Pleiades adopted from Soderblom et al. 
(\cite{Soderblometal93b}). The dashed line shows the upper  envelope for the Hyades cluster 
taken from Thorburn et al. (\cite{Thorburnetal93}).
}
\label{soder}
\end{figure}

The spectroscopic survey resulted in the detection of significant \LiI\ absorption 
lines in a large fraction of
the G and K stars in our sample. In 51 G-K stars lithium absorption lines with an EW larger
than 60\,\AA\ were found. The number of lithium-rich M-type stars is very small. We
found significant \LiI\ absorption lines 
in only 2 out of 47 observed M stars. 
In Fig. \ref{soder} the EWs of \LiI$\lambda6708$\,\AA\  are plotted versus $\teff$ 
for the entire sample.
In Fig. \ref{EWLifield} in the Appendix the same plots are shown for the individual study areas.

\begin{figure}[tb]
\resizebox{\hsize}{!}{\includegraphics[angle=-90,bbllx=80pt,bblly=60pt,bburx=540pt,bbury=790pt,clip=]{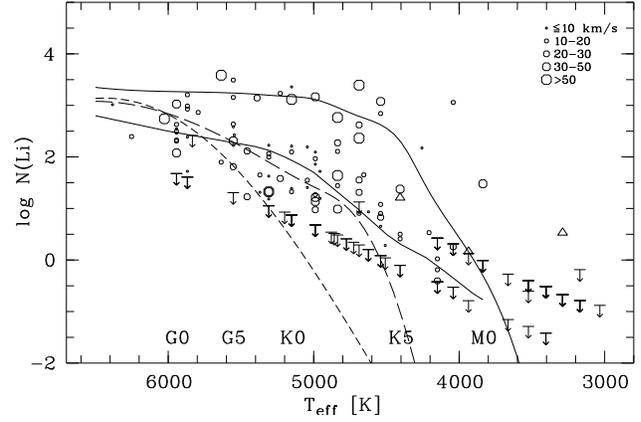}}
\caption[]{Lithium abundances versus effective temperature for the complete sample. 
Upper limits are plotted as
downward arrows. Circles  denote high-resolution measurements with the symbol size
depending on $v\,\sin i$. Low and medium 
resolution data are plotted as triangles. The solid lines are the upper and 
lower limit of $\log N$(\LiI) in the
Pleiades; the long dashed and short dashed lines show the upper $\log N$(\LiI)
limits for the UMaG and the Hyades, respectively. 
}
\label{NLiteff}
\end{figure}

The lithium equivalent widths were converted to abundances, $N$(Li), 
by using the curves of
growth of Soderblom et al. (\cite{Soderblometal93b}) for stars with $\teff > 4000$\,K
and of Pavlenko \& Magazz\`u (\cite{PavlenkoMagazzu96}) and Pavlenko et al. 
(\cite{Pavlenkoetal95}) for cooler stars. As in Paper VI effective temperatures 
were derived from the spectral types using the temperature calibrations of 
de Jager \& Nieuwenhuijzen (\cite{deJagerNieuwenhuijzen87}).
The uncertainty of $\teff$ is typically 200\,K. This leads to errors of the estimated Li 
abundances of about 0.3\,dex. Lithium abundances are shown in 
Fig. \ref{NLiteff} as function of effective temperature with
$v\,\sin~i$ indicated by the symbol size.

\subsubsection{Classification of age groups}
\label{agegroups}
In order to obtain information about the age distribution of the sample stars 
we compared our lithium measurements with the corresponding measurements of 
stars in clusters with various ages: \object{IC\,2602} (30\,Myr), the
\object{Pleiades} (100\,Myr), \object{M\,34} (200\,Myr), the Ursa Major group (UMaG, 300\,Myr), and 
the \object{Hyades} (660\,Myr). Ages are from Lang (\cite{Lang92}) except for IC\,2602 for
which we adopted the age given by Stauffer et al. (\cite{Staufferetal97}). 
The lithium data were taken from Randich et al. (\cite{Randichetal97}) for IC\,2602, 
Soderblom et al. (\cite{Soderblometal93b}) for
the Pleiades,  Jones et al. (\cite{Jonesetal97}) for M\,34,  
Soderblom et al. (\cite{Soderblometal93a}) for UMaG, and Thorburn et al. 
(\cite{Thorburnetal93}) for the Hyades. 
Note that these investigations all use the
same curves of growth by Soderblom et al. (\cite{Soderblometal93b}) for the
conversion of equivalent widths to lithium abundances. 
Upper envelopes of the lithium abundances were
adopted from the cited lithium data. For the Pleiades we also adopted the lower envelope.

In Fig. \ref{soder} the upper and lower envelopes of the 
$\teff - W$(\LiI) distributions for stars in the Pleiades and the upper 
envelope for the Hyades are shown. Likewise, Fig. \ref{NLiteff} 
includes the upper envelopes of the lithium 
abundances of stars in the Pleiades, the UMaG, and the Hyades, and  in addition 
the lower envelope for the Pleiades. 

Using the lithium abundance data for the mentioned clusters and moving groups we
finally  defined four age groups. 
The age group ``PMS'' consists of stars
above the Pleiades upper envelope and is thus younger than the
Pleiades, i.e. younger than 100\,Myr. The group of stars between the upper and lower 
Pleiades envelopes can be assumed to have an age similar to the Pleiades. In the 
Pleiades the G and K stars  are supposed to have reached the ZAMS. This  group with 
an age of $\sim100$\,Myr is therefore designated ``Pl\_ZAMS''. 
The age group ``UMa'' comprises stars between the lower Pleiades and the upper 
Hyades envelope. The age of the stars of this group is between $\sim100$ and 
$\sim600$\,Myr, i.e. on the average $\sim$300\,Myr, which is the age of the UMaG.
The age group ``Hya+'' comprises G-K stars with either a lithium abundance below the 
upper Hyades envelope or with an upper limit for the lithium abundance only. 
The latter means that this group also contains stars for which the upper limit is 
above the Hyades line. Evolved stars more luminous than LC IV are included in the age 
group ``Hya+'' if not stated otherwise in the following. It should be noted, however, 
that due to the well-known scatter of the lithium abundances in clusters stars below the
upper envelope for the corresponding age group are not necessarily older than the
respective group. Therefore, the 
``Hya+'' group might actually also contain some younger stars although it certainly is
dominated by truly old stars. 

In M stars older than several $10^6$\,yr lithium has been destroyed already (e.g. D'Antona \& Mazzitelli
\cite{DAntonaMazzitelli94}). With the 
exception of two stars we could not detect lithium in the M stars of our sample.
This means that the M stars are typically older than $\sim10$\,Myr. 
We thus only defined a group ``M stars'' without assigning an
age. This group does not contain the two lithium rich M stars (see below).
We will return to  the M stars in Sect. \ref{propermotions} where we use the kinematical properties to
estimate their age.

\begin{figure}
\centering
\resizebox{\hsize}{!}{\includegraphics[angle=0,bbllx=65pt,bblly=65pt,bburx=515pt,bbury=740pt,clip=]{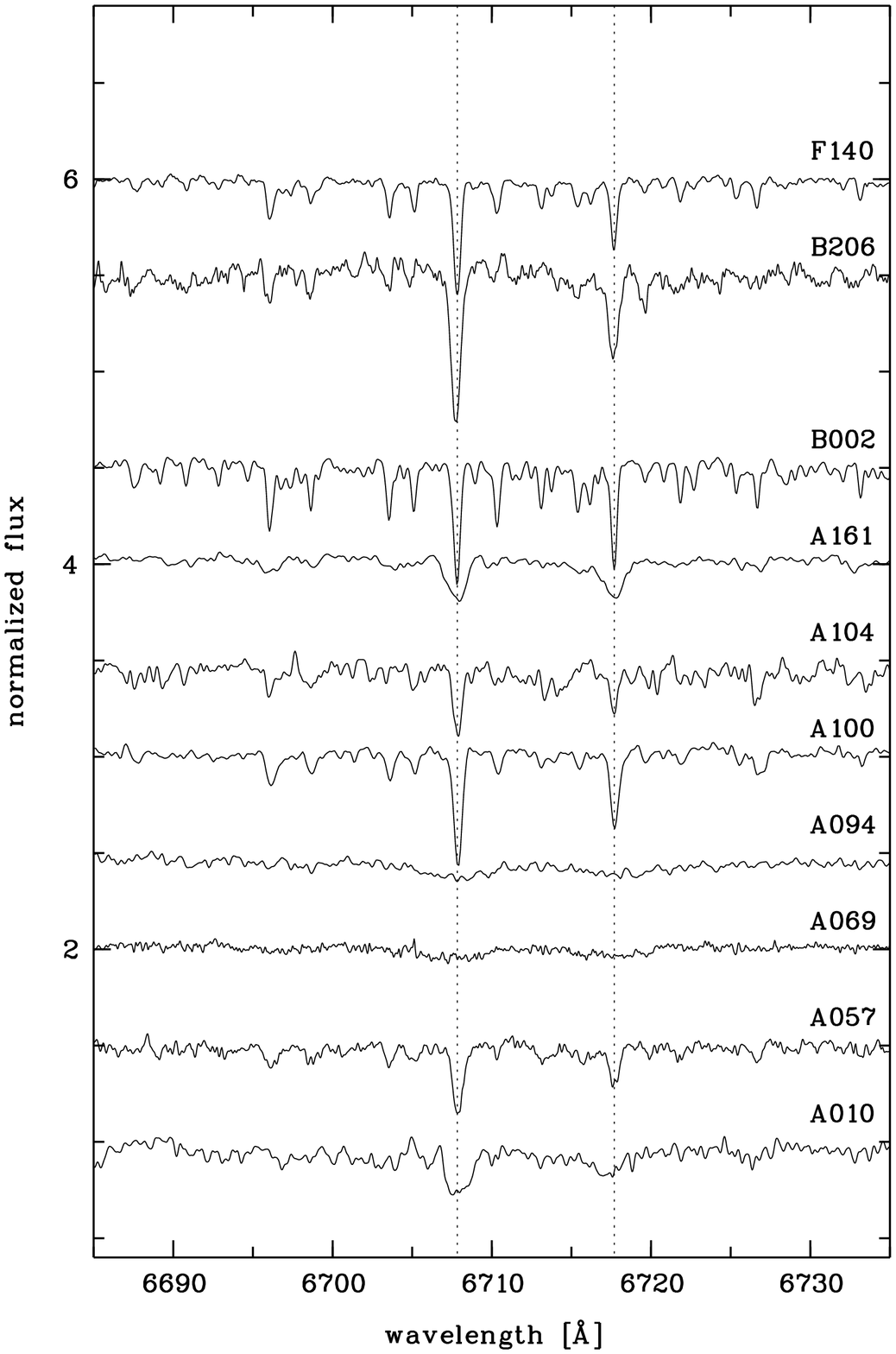}}
\caption[]{Spectra of the lithium-rich sample listed in Table \ref{lirich}.  The wavelengths of
\LiI$\lambda6708$\,\AA\ and \CaI$\lambda6718$\,\AA\ are indicated by the dashed lines.
}
\label{speclirich}
\end{figure}

\begin{figure}
\centering
\resizebox{\hsize}{!}{\includegraphics[angle=-90,bbllx=70pt,bblly=40pt,bburx=550pt,bbury=715pt,clip=]{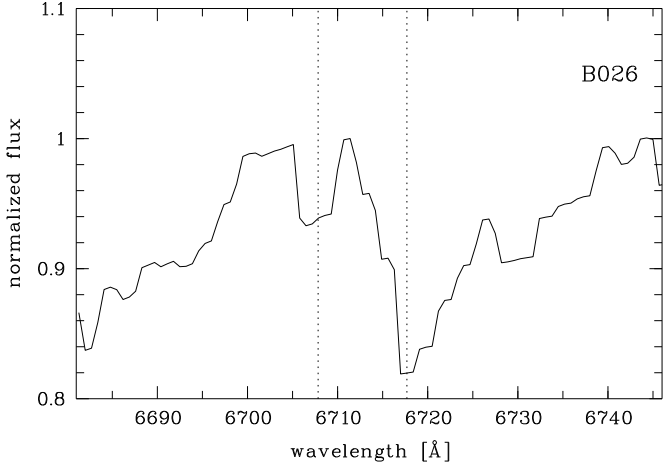}}
\caption[]{Low-resolution spectrum of the M4 star B026. The dashed lines indicate
\LiI$\lambda6708$\,\AA\ and \CaI$\lambda6718$\,\AA .
}
\label{liB026}
\end{figure}

Figure \ref{NLiteff} shows that a small but significant 
group of 12 stars exists above the Pleiades upper limit. These objects appear 
thus to be younger than $\sim100$\,Myr and may be even younger than or comparable to the
age of IC\,2602, i.e. $\sim30$\,Myr. 
Two of these stars, B002 and F0140, are however giants (LC III) and are therefore 
not pre-main sequence (PMS) but evolved objects. 
This leaves a group of 10 stars which appears to consist of PMS  
objects, i.e. true members of the age group ``PMS''.
Actually, 8 of these 10
stars are found in area\,I which is located south of the Tau-Aur SFR. 
They represent the young stellar population in this region discussed in Paper VI. 
The remaining two stars are located in area II.
The subsample of the lithium-rich stars including the giants is listed in Table \ref{lirich}. 
Their high-resolution spectra are shown in Fig. \ref{speclirich} except for A058. 
The spectrum of this star can be found in Neuh\"auser et al. (\cite{Neuhaeuseretal95}). 
For its low-resolution spectrum 
see Paper VI. The spectrum of the M4 star B026 is displayed separately 
in Fig. \ref{liB026}. 

The rotational velocities of the Li-rich stars are high on the average. Only the 
giants have $v\sin i$ below 10\,\kms . Six of the ten PMS stars have 
$v\sin i \ge 20$\,\kms . Table \ref{vrotmed} lists the median $v\sin i$ for each age
group. It shows that $v\sin i$ decreases on the average with increasing age.

\begin{table} 
\centering
\caption[]{Median $v\sin i$ (in \kms ) for the different age groups. Giants  
were not included in group Hya+.
} 
\begin{tabular}{lllll} 
\noalign{\smallskip}     
\hline 
\hline 
\noalign{\smallskip}    
age group   & PMS &  Pl\_ZAMS & UMa & Hya+ \\
\noalign{\smallskip}    
\hline 
\noalign{\smallskip}    
${\langle v\sin i \rangle}_{\rm med}$   &  32  &  17 & 18 & 11\\ 
\noalign{\smallskip}    
\hline 
\noalign{\smallskip}    
\label{vrotmed}
\end{tabular}
\end{table}

\begin{table} 
\tabcolsep=3pt
\centering
\caption[]{Spectral types, lithium equivalent widths, $EW$(Li),  
logarithmic abundances, $\log N$(Li), and projected rotational velocities for 
the subsample of stars with lithium abundance above the Pleiades upper envelope. 
(``PMS'' sample). Evolved 
lithium-rich stars  not belonging to the PMS sample are marked by the ``${\ast}$'' symbol.
} 
\centering
\begin{tabular}{llllll} 
\noalign{\smallskip}     
\hline 
\hline 
\noalign{\smallskip}    
field & RASS name        &Sp. type&  $EW$(\LiI )   & $\log N$(Li) & $v\,\sin i$  \\            
      &  &  &[m\AA ] &   &  [\kms ]  \\
\noalign{\smallskip}    
\hline 
\noalign{\smallskip}    
A010            & RXJ0331.1$+$0713   & K4Ve       & $ $ 407&$   3.08 $&$   42$ \\ 
A057            & RXJ0344.4$-$0123   & G9V-IV     & $ $ 277&$   3.23 $&$   20$ \\ 
A058            & RXJ0344.8$+$0359   & K1Ve       & $ $ 310&$   3.17 $&$   31$ \\ 
A069            & RXJ0348.5$+$0831   & G4V:       & $ $ 259&$   3.59 $&$> 100$ \\ 
A094            & RXJ0355.2$+$0329   & K3V        & $ $ 424&$   3.39 $&$> 100$ \\ 
A100            & RXJ0358.1$-$0121   & K4V        & $ $ 362&$   2.84 $&$   15$ \\ 
A104            & RXJ0400.1$+$0818   & G5V-IV     & $ $ 259&$   3.49 $&$   12$ \\ 
A161            & RXJ0417.8$+$0011   & M0Ve       & $ $ 304&$   1.48 $&$   44$ \\ 
B002$^{\ast}$        & RXJ0638.9$+$6409   & K3III      & $ $ 315&$   2.17 $&$    6$ \\ 
B026            & RXJ0708.7$+$6135   & M4e        & $ $ 272&$   0.53 $&$    *$ \\ 
B206            & RXJ0828.1$+$6432   & K8Ve       & $ $ 607&$   3.06 $&$   16$ \\ 
F140$^{\ast}$        & RXJ2241.9$+$1431   & K0III      & $ $ 307&$   3.36 $&$<   5$ \\ 
\hline 
\noalign{\smallskip}    
\end{tabular}
\label{lirich}
\end{table}

\begin{table} 
\tabcolsep=3pt
\centering
\caption[]{Spectral types, lithium equivalent widths, $EW$(Li),  
logarithmic abundances, $\log N$(Li), and projected rotational velocities for 
the subsample of stars with lithium abundance between the Pleiades lower and 
upper envelope (``Pl\_ZAMS'' sample). Lithium-rich evolved stars not belonging to the 
Pl\_ZAMS sample are marked by ``${\ast}$''.
} 
\centering
\begin{tabular}{llllll} 
\noalign{\smallskip}     
\hline 
\hline 
\noalign{\smallskip}    
field & RASS name        &Sp. type&  $EW$(\LiI )   & $\log N$(Li) & $v\,\sin i$  \\            
      &  &  &[m\AA ] &   &  [\kms ]  \\
\noalign{\smallskip}    
\hline 
\noalign{\smallskip}    
A001            & RXJ0328.2$+$0409   & K0         & $ $ 275&$   3.12 $&$   96$ \\ 
A036            & RXJ0338.7$+$0136   & K4Ve       & $ $  80&$   1.08 $&$   17$ \\ 
A039            & RXJ0338.8$+$0216   & K4         & $ $  58&$   0.90 $&$   16$ \\ 
A042            & RXJ0339.9$+$0314   & K2         & $ $ 104&$   1.64 $&$   63$ \\ 
A056            & RXJ0343.9$+$0327   & K1V-IV     & $ $ 126&$   1.97 $&$   12$ \\ 
A063            & RXJ0347.1$-$0052   & K3V        & $ $  84&$   1.31 $&$   21$ \\ 
A071            & RXJ0348.9$+$0110   & K3V:e      & $ $ 258&$   2.36 $&$   83$ \\ 
A090            & RXJ0354.3$+$0535   & G0V        & $ $ 131&$   3.03 $&$   31$ \\ 
A095            & RXJ0355.3$-$0143   & G5V        & $ $ 213&$   3.16 $&$   19$ \\ 
A096            & RXJ0356.8$-$0034   & K3V        & $ $ 122&$   1.55 $&$   22$ \\ 
A101            & RXJ0358.9$-$0017   & K3V        & $ $ 294&$   2.62 $&$   27$ \\ 
A120            & RXJ0404.4$+$0518   & G7V        & $ $ 239&$   3.15 $&$   27$ \\ 
A126            & RXJ0405.6$+$0341   & G0V-IV     & $ $  63&$   2.54 $&$<   5$ \\ 
A154            & RXJ0416.2$+$0709   & G0V        & $ $  58&$   2.50 $&$   11$ \\ 
B008            & RXJ0648.5$+$6639   & G5         & $ $ 121&$   2.58 $&$    9$ \\ 
B018            & RXJ0704.0$+$6214   & K5Ve       & $ $  36&$   0.50 $&$   12$ \\ 
B034$^{\ast}$        & RXJ0714.8$+$6208   & G1IV-III   & $ $ 127&$   2.93 $&$   12$ \\ 
B039            & RXJ0717.4$+$6603   & K2V        & $ $ 213&$   2.27 $&$   21$ \\ 
B068            & RXJ0732.3$+$6441   & K5e        & $ $ 130&$   1.21 $&$    *$ \\ 
B086$^{\ast}$        & RXJ0742.8$+$6109   & K0III      & $ $ 147&$   1.65 $&$   11$ \\ 
B124$^{\ast}$        & RXJ0755.8$+$6509   & G5III      & $ $ 153&$   2.19 $&$    6$ \\ 
B160            & RXJ0809.2$+$6639   & G2V        & $ $ 128&$   2.86 $&$   12$ \\ 
B174            & RXJ0814.5$+$6256   & G1V        & $ $ 136&$   2.99 $&$   11$ \\ 
B183            & RXJ0818.3$+$5923   & K0V        & $ $ 134&$   2.21 $&$    7$ \\ 
B185            & RXJ0819.1$+$6842   & K7Ve       & $ $  26&$   0.02 $&$   13$ \\ 
B199            & RXJ0824.5$+$6453   & K4V        & $ $  50&$   0.83 $&$   26$ \\ 
C047            & RXJ1027.0$+$0048   & G0V        & $ $  73&$   2.63 $&$   11$ \\ 
C058            & RXJ1028.6$-$0127   & K5e        & $ $  30&$   0.41 $&$   12$ \\ 
C143            & RXJ1051.3$-$0734   & K2V        & $ $  75&$   1.44 $&$   18$ \\ 
C165            & RXJ1057.1$-$0101   & K4V        & $ $  34&$   0.65 $&$    7$ \\ 
C176            & RXJ1059.7$-$0522   & K1V        & $ $ 148&$   2.09 $&$    9$ \\ 
C197            & RXJ1104.6$-$0413   & G5V        & $ $ 130&$   2.64 $&$   10$ \\ 
C200            & RXJ1105.3$-$0735   & K5e        & $ $ 160&$   1.38 $&$   32$ \\ 
D064            & RXJ1210.6$+$3732   & K0         & $ $ 115&$   2.10 $&$   19$ \\ 
E022            & RXJ1628.4$+$7401   & G1V        & $ $ 172&$   3.21 $&$   18$ \\ 
E067            & RXJ1653.5$+$7344   & G1IV       & $ $  98&$   2.43 $&$    8$ \\ 
E179            & RXJ1728.1$+$7239   & K4IVe      & $ $  66&$   0.53 $&$   18$ \\ 
F015            & RXJ2156.4$+$0516   & K2         & $ $ 185&$   2.11 $&$   12$ \\ 
F046            & RXJ2212.2$+$1329   & G8:V:      & $ $ 106&$   2.23 $&$    5$ \\ 
F060            & RXJ2217.4$+$0606   & K1e        & $ $ 108&$   1.86 $&$    8$ \\ 
F087            & RXJ2226.3$+$0351   & G5:V:      & $ $  80&$   2.31 $&$   31$ \\ 
F101            & RXJ2232.9$+$1040   & K2V:       & $ $ 285&$   2.76 $&$   97$ \\ 
F142            & RXJ2242.0$+$0946   & K8V        & $ $  54&$   0.26 $&$   24$ \\ 
\hline 
\noalign{\smallskip}    
\end{tabular}
\label{zams}
\end{table}

The majority of stars has EWs and lithium abundances below the Pleiades 
upper limits of EW and $\log N$(Li), respectively. In the region between 
the upper and lower envelope of the Pleiades  43 G-K stars are found.  
This group is listed in Table \ref{zams}. Three of these stars 
are giants with LC IV-III, III,  and II.
The 40 non-giants appear to constitute a population with an age similar to the 
Pleiades, i.e. $\sim$100\,Myr. 
The region between the Hyades upper and the Pleiades lower envelope 
contains 23 stars of which 4 are evolved objects.
The UMa age group with an age of $\sim$300\,Myr thus consists of 19 stars. 
Below the upper limit of the Hyades 57 non-giant stars are found and are 
thus assigned an age of older than $\sim$600-700\,Myr. Adding the 17 evolved
G-K stars which are certainly also older than $\sim$1\,Gyr results in a total 
of 74 stars for age group ``Hya+''. 
Thus lithium abundances and luminosity classification suggest that 47\% of all G-K stars 
in the sample have an age of less than about 600-700\,Myr. 
Restricting these statistical considerations to the later spectral types 
increases the fraction of stars younger than the Hyades. Of the 114 G5-K9 stars 55, 
i.e. $\sim50$\%, have a lithium abundance higher than the 
Hyades. With the above mentioned ambiguity of the age group definition 
this means that at least half of the G5-K9 stars are
younger than the Hyades. Some statistics of the age distribution of our sample
stars for these age groups is summarized in Table \ref{age}.

\subsubsection{Spatial distribution of the age groups}
The spatial distribution of the G and K stars of the various age groups is summarized in 
Table \ref{agelocation}. Variations of the surface density of the various age groups
with location are indicated. 

As expected  area IV located near the north galactic pole has 
the lowest surface density of stars younger than the Hyades. In this area only 2 
stars younger than $600-700$\,Myr are found in 72\,deg$^{2}$. This 
corresponds to a surface density of 
$0.028\pm0.020$\,deg$^{-2}$ 
at a RASS count-rate limit of 0.03\,cts\,s$^{-1}$. 
In the other 5 areas (613.2\,deg$^{2}$) a total of 60 stars (including 5 
stars in area V above 0.03\,cts\,s$^{-1}$) yields a surface density of 
$0.0978\pm0.013$\,deg$^{-2}$. 
Counting stars of all age groups area IV has a surface density of 
$0.097\pm0.037$\,deg$^{-2}$ 
compared to 
$0.204\pm0.018$\,deg$^{-2}$
in the other areas at the same count-rate limit.
A t-test shows that these differences are significant. 

The very young stars of the PMS sample are apparently more abundant in area I 
than in any other area: 80\% of these stars are found in area I. 
Adding up the  numbers of stars younger than the Hyades in areas II, III, and VI 
leads to an average surface density 
$0.077\pm0.013$\,deg$^{-2}$. 
This is less than half
of the value in area I which is  
$0.167\pm0.034$\,deg$^{-2}$ . 
Although indicative for a higher concentration of young stars in
area\,I the difference is not significant.

\begin{table} 
\tabcolsep=3pt
\centering
\caption[]{Statistics of the age distribution of the sample of G-K stars. ``PMS'' denotes stars
younger than 100\,Myr, ``Pl\_ZAMS'' stars as old as the Pleiades, ``UMa'' stars 
with an age of $\sim300$\,Myr, and ``Hya+'' older than the Hyades. The latter age group also contains 
17 evolved stars (LC IV-III, III, and II). The total number of G-K stars is 141.
} 
\begin{tabular}{lllll} 
\noalign{\smallskip}     
\hline 
\hline 
\noalign{\smallskip}    
       & \multicolumn{4}{c}{age group} \\ 
       & PMS & Pl\_ZAMS & UMa  &  Hya+ \\ 
       & $<100$\,Myr & 100\,Myr & $\sim300$\,Myr & $>660$\,Myr \\ 
\hline 
\noalign{\smallskip}    
number  G-K& 8 & 40  & 19  & 74 \\ 
fraction G-K& 6\% & 28\%  & 13\%  & 52\% \\ 
\noalign{\smallskip}    
\hline 
\noalign{\smallskip}    
\end{tabular}
\label{age}
\end{table}

\begin{table} 
\tabcolsep=5pt
\centering
\caption[]{Statistics of the spatial distribution of the various age groups 
in the sample. For each age group the total number of stars and the number per square
degree is given. The numbers are for a RASS count-rate limit of 0.03\,cts\,s$^{-1}$
except for area V which has a count-rate limit of 0.01\,cts\,s$^{-1}$.
} 
\begin{tabular}{lllllllll} 
\noalign{\smallskip}     
\hline 
\hline 
\noalign{\smallskip}    
area        & \multicolumn{8}{c}{age group} \\ 
\noalign{\smallskip}    
\hline 
\noalign{\smallskip}    
       & \multicolumn{2}{c}{PMS} & \multicolumn{2}{c}{Pl\_ZAMS} &\multicolumn{2}{c}{UMa}  &  \multicolumn{2}{c}{Hya+} \\ 
       & \multicolumn{2}{c}{$<100$\,Myr} & \multicolumn{2}{c}{100\,Myr} & \multicolumn{2}{c}{$\sim300$\,Myr} & \multicolumn{2}{c}{$>660$\,Myr} \\ 
\noalign{\smallskip}    
\hline 
\noalign{\smallskip}    
I     & 8 & 0.056 & 14 & 0.097 &  3  & 0.021 & 10 & 0.069 \\  
II    & 2 & 0.014 & 9  & 0.063 &  1  & 0.007 & 22 & 0.153 \\ 
III   & 0 & 0     & 7  & 0.049 &  1  & 0.007 & 16 & 0.111 \\ 
IV    & 0 & 0     & 1  & 0.014 &  1  & 0.014 &  5 & 0.069 \\ 
V     & 0 & 0     & 3  & 0.084 &  6  & 0.161 & 11 & 0.296 \\ 
VI    & 0 & 0     & 6  & 0.042 &  7  & 0.049 & 10 & 0.069 \\ 
\noalign{\smallskip}    
\hline 
\noalign{\smallskip}    
\end{tabular}
\label{agelocation}
\end{table}

\subsubsection{Age dependent $\log N - \log S$ distribution}
We  compared the observed cumulative number distribution, $\log N(>S) - \log S$,
of our sample with model predictions by Guillout et al. (\cite{Guilloutetal96}).  
The median latitude for the combined areas I, II, III, V, and VI, which are distributed 
between galactic latitudes of 20$\degr$ and 50$\degr$, is actually $30\degr$, thus matching this model
parameter well. The models of Guillout et al. (\cite{Guilloutetal96}) give cumulative surface 
densities, $N(>S)$, as a function of ROSAT-PSPC count rate, $S$, for three age bins:  age younger
then 150\,Myr, age between 150\,Myr and 1\,Gyr, and older than 1\,Gyr. We  restricted
the comparison to the youngest model age bin and to the sum
of all model age bins because of the difficulty to separate observationally 
stars with ages of several 100\,Myr to $\sim1$\,Gyr and older.
We further considered the combined sample of G and K stars. M stars were not included
because of the lack of an observational age determination for stars of this spectral type in
our sample.
The uncertainty of the ages derived observationally from lithium was taken into account by forming two
observational age samples matching as closely as possible the youngest age bin of 
the models: a) a sample comprising the sum of G-K stars from the PMS and Pl\_ZAMS age
group, and b) a sample containing in addition the corresponding UMa stars. The true
sample of stars younger than 150\,Myr is expected to lie between these limits.

\begin{figure}
\centering
\resizebox{\hsize}{!}{\includegraphics[angle=-90,bbllx=75pt,bblly=50pt,bburx=540pt,bbury=770pt,clip=]{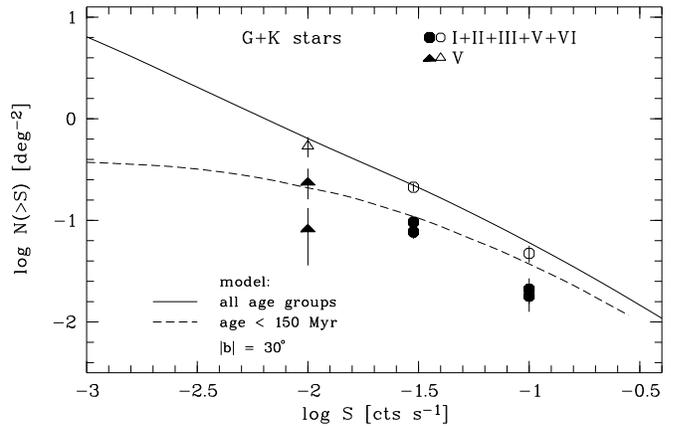}}
\caption[]{Comparison of observed co-added number densities of G and K stars, $N(>S)$, for three RASS count 
rates $S$ with models of Guillout et al. (\cite{Guilloutetal96}) for $|b| = 30\degr$. 
Open symbols denote the sum of all age groups. Lower and upper filled symbols 
represent  the sum of age groups ``PMS'' and
``Pl\_ZAMS'', and of ``PMS'', ``Pl\_ZAMS'', and UMa, respectively. 
}
\label{guillout}
\end{figure}

The result of the comparison of $\log N(>S) - \log S$ is depicted in Fig. 
\ref{guillout} for three RASS X-ray count rates of
0.1, 0.3 and 0.01\,cts\,s$^{-1}$. The predicted  numbers of G-K are in good agreement 
with our sample in the 5 study areas located around $30\degr$ 
in galactic latitude. This holds for both the sum of all age groups and stars younger than
$\sim150$\,Myr obtained as described above and represented in the figure 
by the filled symbols. Likewise, the predicted flattening of $\log N(>S) - \log S$ 
at lower count rates is also found in our data for area V which has the lowest
count rate limit of 0.01\,cts\,s$^{-1}$. 

\subsection{Kinematics}
\subsubsection{Proper motions}
\label{propermotions}
\begin{figure}
\resizebox{\hsize}{!}{\includegraphics[angle=0,bbllx=70pt,bblly=15pt,bburx=580pt,bbury=720pt,clip=]{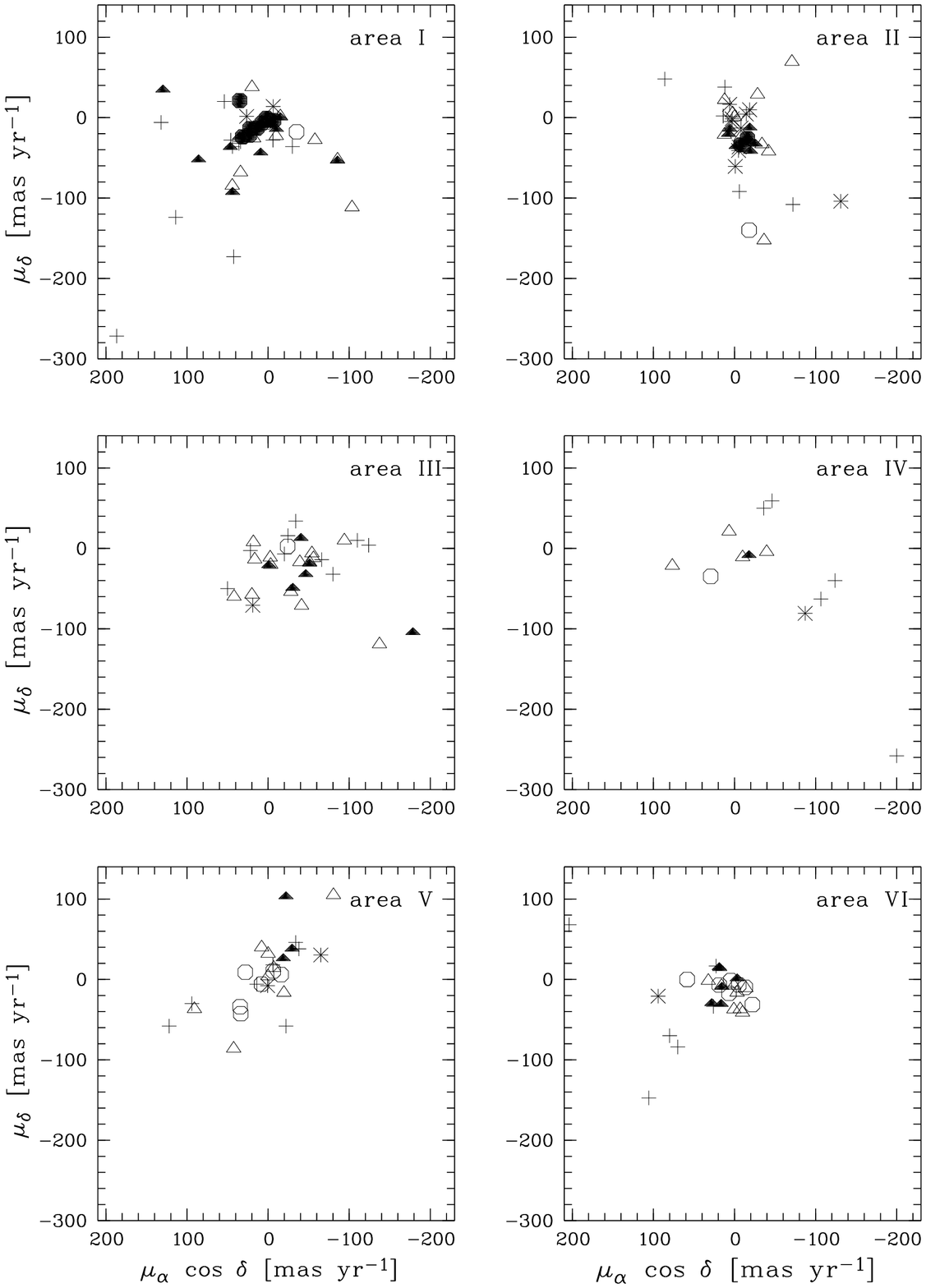}}
\caption[]{Proper motions for the six study areas. 
The different symbols denote the different age groups: filled circles = PMS, filled 
triangles =  Pl\_ZAMS, open circles = UMa, open triangles = Hya+, $\ast$ = giants, $+$ = M
stars without Li detection. 
}
\label{properall}
\end{figure}
We searched for proper motions in a variety of different catalogs:
the Hipparcos Catalog (ESA \cite{Hipparcos}),  
the Positions and Proper Motions Catalog (PPM) (R\"oser \& Bastian \cite{PPM}),  
the ACT Reference Catalog (Urban et al. \cite{ACT}), the
Tycho Reference Catalogue (TRC)  (Hog et al. \cite{Hogetal98}), 
the Tycho-2 catalog (Hog et al. \cite{Hogetal00}), the STARNET catalog
(R\"oser \cite{STARNET}), and 
the Second U.S. Naval Observatory CCD Astrograph Catalog (UCAC2) 
(Zacharias et al. \cite{UCAC2}). The PPM and the STARNET catalogs
were locally transformed to the Hipparcos reference system before identification.
For many stars we found entries
in more than one catalog, and in these cases the proper motions were compared
and the one which had consistent solutions across several catalogs was usually
chosen. If all proper motions were consistent, the most precise one was
adopted; this was usually the Hipparcos or the UCAC2 proper motion (the
Hipparcos catalog has a high weight in the solution for the UCAC2 proper
motion), or the Tycho-2 proper motion for those regions not covered yet by 
the UCAC2 catalog. However, in many cases the
proper motion in Hipparcos differed from the entries in other catalogs,
which is likely due to the fact that the Hipparcos proper motions reflect
the 'instantaneous' motion during the Hipparcos mission, which is often
affected by orbital motion, whereas most of the proper motions in the other
catalogs are based on observations stretched out over a longer baseline
and thus better reflect the real motion of the center of mass through space
which is of interest here.

Altogether, we were able to assign proper motions to the counterparts of 129 
RASS sources with spectral types G to M. 
In detail we found  55 of 56 G stars, 61 of 86 K stars and 13 
of 56 M stars in the mentioned catalogs. In addition we also found 54 F stars.
An equal number of proper motions comes from Tycho-2 and UCAC2, while only two
proper motions each were taken from Hipparcos and TRC, and only one each
from PPM and STARNET, while the ACT was not used in the end at all.

These proper motion data were supplemented for the optically faint stars (mainly of spectral
type K and M) by data from other catalogs: 53 stars from USNO-B1.0 
(Monet et al. \cite{USNOB1}, 36 M stars, 16 K stars, and 1 G star),
1 M star from Carlsberg Meridian Catalogs (\cite{Carlsberg}), and 1 M star 
from the NPM1 Catalog of the Lick Northern Proper Motion Program  
(Klemola et al. \cite{Klemolaetal87}). Note that 
the USNO-B1.0 proper motions are not absolute, but relative to the Yellow Sky 
Catalog YS4.0 in the sense that the mean motion of objects common to USNO-B1.0 
and YS4.0 was set to zero in USNO-B1.0. 
According to  Monet et al. (\cite{USNOB1}) 
the difference between these relative proper motions and the true absolute
ones should, however, be small.

Thus in total proper motion data are available for all G stars, 
for 77 of 86 K stars, and for 51 of 56 M stars.

The proper motions are shown in Fig. \ref{properall} for the individual study areas. 
The diagram displays proper motions for six groups of stars, i.e. the ``PMS''
''Pl\_ZAMS'', ``UMa'' and ``Hya+'' age groups, evolved stars (giants) and 
the M stars without lithium detection.

Of particular interest are the proper motions of the stars of the youngest age groups in area\,I, 
i.e. the ``Pl\_ZAMS'' and ``PMS'' samples with ages of 100\,Myr and less,   
This study area is located near the Tau-Aur SFR and near the Gould Belt 
(see Fig. \ref{area_gal}) and has, probably due to its
location,  the highest surface density of young stars.
Proper motions exist for all eight young stars of area\,I listed in Table \ref{lirich}. 
They are plotted in the upper left panel of  Fig. \ref{properall}.
Four of the stars of age group PMS in area\,I were already identified as pre-main sequence objects 
by Neuh\"auser et al. (\cite{Neuhaeuseretal97}) (A058, A069, A090, and A104). They 
assigned an age of 35\,Myr to these stars.
Likewise, one star of the Pl\_ZAMS sample, A120, was assigned an age of 100\,Myr
by Neuh\"auser et al.. These stars were part of a sample investigated kinematically for
membership to the Taurus-Auriga SFR by Frink et al. (\cite{Frinketal97}).  They 
studied stars in the central region of Tau-Aur and in a region south of Tau-Aur 
which partially overlaps at the southern edge with our area\,I. 
The sample studied by
these authors contains three further stars of our sample, A007, A107, and A122, for which 
Neuh\"auser et al. assigned an age 
of older than 100\,Myr. This is in agreement with our age estimate of older 
than 660\,Myr for A007 and A107, and of $\sim$300\,Myr for A122. 

In Table \ref{pmarea1} the mean proper motions and their dispersions are 
summarized for the PMS, Pl\_ZAMS, UMa, and Hya+ age groups, and for M stars 
without \LiI\ detection.
Obviously,  the 8 ``PMS'' stars  show a smaller spread in proper motions than the older stars.
They cluster around ($\mu_{\alpha}\cdot \cos\delta$, $\mu_{\delta}$) of ($+16,
-8$)\,mas\,yr$^{-1}$ with a scatter of $\sim15$\,mas\,yr$^{-1}$ in each direction. 
Frink (\cite{Frink99}) 
transformed the proper motions given by Frink et al. (\cite{Frinketal97}) from the 
FK5 to the Hipparcos system and determined mean values of 
($+8.7,-11.2$)\,mas\,yr$^{-1}$ for the southern  sample of Frink et al. (\cite{Frinketal97}). 
For the central region of Tau-Aur Frink (\cite{Frink99}) derived mean proper motions 
of ($+4.5,-19.7$)\,mas\,yr$^{-1}$. The comparison of our results with the findings of Frink 
(\cite{Frink99}) reveals an interesting trend in the mean proper motions 
{\em relative} to the core region of Tau-Aur. The southern sample of Frink et al. moves away
from the centre of Tau-Aur with a mean proper motion of (+4.2,+8.5)\,mas\,yr$^{-1}$. The PMS
stars in area I are located even more to the south of the centre and their relative 
mean proper motion is actually even larger, (+12,+12)\,mas\,yr$^{-1}$. Thus we find that 
the stars in area I move in approximately the same direction as the southern stars of 
Frink et al., but with an even higher proper motion. 

Inspection of Fig. 2 in Frink et al. (\cite{Frinketal97}) allows to
estimate a dispersion of about 15 to 20\,mas\,yr$^{-1}$ for both subsamples 
which again is compatible with the 15\,mas\,yr$^{-1}$ derived for our PMS subsample.
The Pl\_ZAMS  stars exhibit a dispersion of the  
proper motion which is larger
by a factor of 2 to 3. On the other hand, the UMa sample though being older 
shows  more coherent proper motions with a dispersion equal to the PMS stars.
The old stars of the Hya+ group and the M stars exhibit the largest dispersions.
Similar results are found for the other study areas.

\begin{table} 
\centering
\caption[]{Mean proper motions and dispersions in area I for  stars  of age groups PMS, 
Pl\_ZAMS, UMa, and M stars without lithium detection (in  mas\,yr$^{-1}$).
} 
\begin{tabular}{lllll} 
\noalign{\smallskip}     
\hline 
\hline 
\noalign{\smallskip}    
age group    &  $\langle\mu_{\alpha}\cos\delta\rangle$ & $\sigma_{\alpha}$ &$ \langle\mu_{\delta}\rangle$ &
    $\sigma_{\delta}$\\
\noalign{\smallskip}    
\hline 
\noalign{\smallskip}    
PMS         & $+16$ & 15 & $-8$ & 14\\
Pl\_ZAMS    & $+18$ & 52 & $-22$ & 32\\
UMa & $+16$ & 15 & $-8$ & 14\\
Hya+ & $-33$ & 69 & $-42$ & 55\\
M stars     & $+59$ & 64 & $-67$ & 131\\
\noalign{\smallskip}    
\hline 
\noalign{\smallskip}    
\end{tabular}
\label{pmarea1}
\end{table}

So far we have considered the proper motions which depend on the distance and contain a
contribution due to the solar motion. We therefore calculated 
tangential velocity components, $v_l$ and $v_b$, in galactic coordinates, $l$ and $b$, by 
using the distance estimates discussed above and the relations 
$v_l = 4.74\times\mu_l\,\cos\,b\times d$\,\kms\ 
and  $v_b = 4.74\times\mu_b\times d$\,\kms , with  
$\mu_l\,\cos\,b$ and $\mu_b$ being proper motions in galactic coordinates given 
in arcsec\,yr$^{-1}$ and the distance $d$ in pc. 
A table summarizing the resulting velocities
and their dispersions for the individual study areas can be found in the Appendix 
(Table \ref{vtangent}). 

The direction-dependent part of the tangential velocities due to the solar reflex motion 
can finally be removed by transforming these velocities to the local standard
of rest (LSR). This is achieved by adding the corresponding solar velocity components. 
We used  the solar motion vector of Dehnen \& Binney 
(\cite{DehnenBinney98}), ($U_{\sun}$, $V_{\sun}$, $W_{\sun}$) = (+10.0, +5.25, +7.17)\,\kms\ (see 
below for the definition of the space velocities) to determine the solar reflex motion:

\begin{equation}
v_{l,\sun} = -U_{\sun}\sin l +  V_{\sun}\cos l 
\label{vsunl}
\end{equation} 
\begin{equation}
v_{b,\sun} = U_{\sun}\cos l \sin b -  V_{\sun}\sin l \sin b + W_{\sun} \cos b
\label{vsunb}
\end{equation}

In contrast to the observed proper motions the tangential velocity components of the  different 
object groups exhibit a similar scatter around the mean of the respective sample. 
This is particularly evident for the M stars which have on average the smallest distances and 
hence have the largest proper motions. 
Generally, the dispersions of their tangential velocities are of the same order of magnitude as
for the other object groups, although there are some differences between the individual study areas.
From the kinematical point of view the M stars in area I appear to be young, $\sim100$\,Myr, 
as they resemble the Pl\_ZAMS group with regard to both the mean velocity and the velocity dispersion.
This also holds for  area III and VI where the M stars kinematically appear somewhat 
older, $\sim300$\,Myr, with velocity dispersions between the Pl\_ZAMS and the Hya+ group.
In area II,  IV,  and V, on the other hand, the M stars show kinematical resemblance to the Hya+ age 
group suggesting an age of $\ga600$\,Myr.

In Table \ref{mugallsr} mean proper motions in galactic coordinates with respect 
to the LSR, $\langle \mu_l \cos b \rangle_{\rm LSR}$ and $\langle \mu_ b \rangle_{\rm LSR}$, and
the corresponding tangential velocities,  $\langle v_l \rangle_{\rm LSR}$ and 
$\langle v_b \rangle_{\rm LSR}$ are listed for the different age groups. As discussed before the M stars
exhibit the largest dispersion of the proper motions. Taking the distance effect into account  the 
dispersions of the respective tangential velocities are reduced to   values similar to those obtained 
for the Pl\_ZAMS and UMa age groups. This again leads to the conclusion that the M stars have 
on the average an age of $\sim$100-600\,Myr. The largest velocity dispersions  are found 
for the Hya+ age group.

\begin{table*} 
\centering
\caption[]{Mean proper motions (in mas\,s$^{-1}$), and mean tangential velocities (in \kms ) 
in galactic coordinates, both  with 
dispersions,  reduced to the LSR. The values are listed  
for stars of the age groups  PMS, Pl\_ZAMS, UMa and Hya+ (split into dwarfs and giants),
and M stars without lithium detection. 
} 
\begin{tabular}{lllll} 
\noalign{\smallskip}     
\hline 
\hline 
\noalign{\smallskip}    
 age group  & $\langle \mu_l\ cos b \rangle_{\rm LSR}$ & $\langle \mu_ b \rangle_{\rm LSR}$ & $\langle
             v_l \rangle_{\rm LSR}$ & $\langle v_b \rangle_{\rm LSR}$\\
\noalign{\smallskip}    
\hline 
\noalign{\smallskip}    
PMS         &$+3 \pm 14$  &$+11 \pm 17$ &$+0 \pm 8$   & $+14 \pm 11$\\
Pl\_ZAMS    &$+8 \pm 29$  &$+2 \pm 35$  &$+2 \pm 18$  & $+5 \pm 14$\\
UMa         &$-8 \pm 37$  &$+0 \pm 27$  &$-14 \pm 36$ &$+0 \pm 15$\\
Hya+:        & & & & \\
\multicolumn{1}{r}{dwarfs}&$-7 \pm 75$  &$-7 \pm 42$ &$-47 \pm 310$&$-26 \pm 184$\\
\multicolumn{1}{r}{giants }     &$+21 \pm 39$ &$-5 \pm 31$  &$+5 \pm 72$&$+4 \pm 34$ \\
M stars     &$+22 \pm 108$&$-34 \pm 139$&$-5 \pm 20$&$-2 \pm 23$ \\
\noalign{\smallskip}    
\hline 
\noalign{\smallskip}    
\end{tabular}
\label{mugallsr}
\end{table*}

\subsubsection{Space velocities}
\label{spacevelocities}
For stars with a distance estimate from trigonometric, spectroscopic or IR photometric
parallax radial velocities and proper motions were combined 
in order to determine the galactic space velocity components $U$, $V$, and $W$. A 
right-handed coordinate system was used with the $U$ axis pointing towards the galactic centre,  
the $V$ axis in the direction of galactic rotation, and the  $W$ axis  towards the north galactic pole. 
The transformation to the LSR was  performed by using the solar motion vector of Dehnen \& Binney 
(\cite{DehnenBinney98}) given above. The required data, RVs, proper
motions, and distances, were available for 44 of 56 G stars, 46 of 85 K stars, and 
7 of 56 M stars.  The space velocity components and related errors were calculated using the
formulae given by Johnson \& Soderblom (\cite{JohnsonSoderblom87}). For the calculation 
of the errors  an uncertainty of 50\% was adopted for the distance for stars with a spectroscopic or 
photometric parallax. The resulting velocity components are listed in Table \ref{resulttab2}.

\begin{figure}
\centering
\resizebox{\hsize}{!}{\includegraphics[angle=0,bbllx=180pt,bblly=130pt,bburx=420pt,bbury=715pt,clip=]{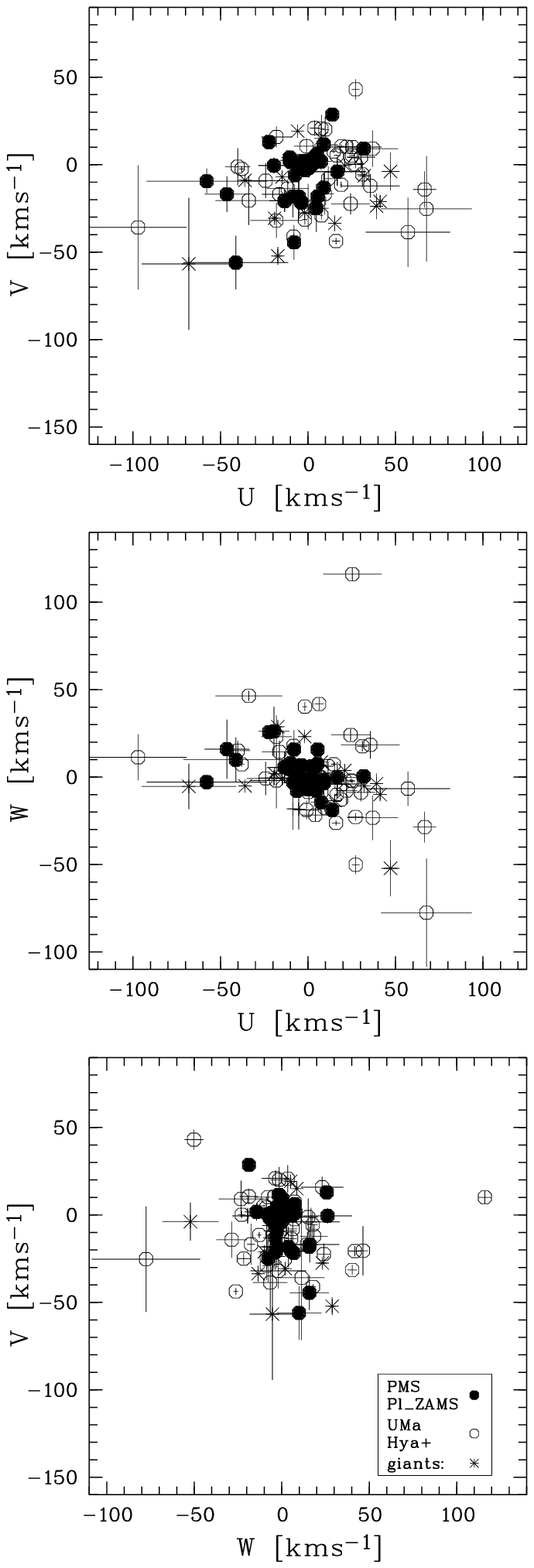}}
\caption[]{Space velocities $U$, $V$, and  $W$  in the LSR frame.
Stars of age groups ``PMS'' and ``Pl\_ZAMS'' are plotted as
filled circles. Open circles denote stars of the UMa and ``Hya+'' age group. 
Giants are plotted as asterisks.
}
\label{UVW}
\end{figure}

\begin{figure}
\centering
\resizebox{\hsize}{!}{\includegraphics[angle=0,bbllx=135pt,bblly=4pt,bburx=515pt,bbury=725pt,clip=]{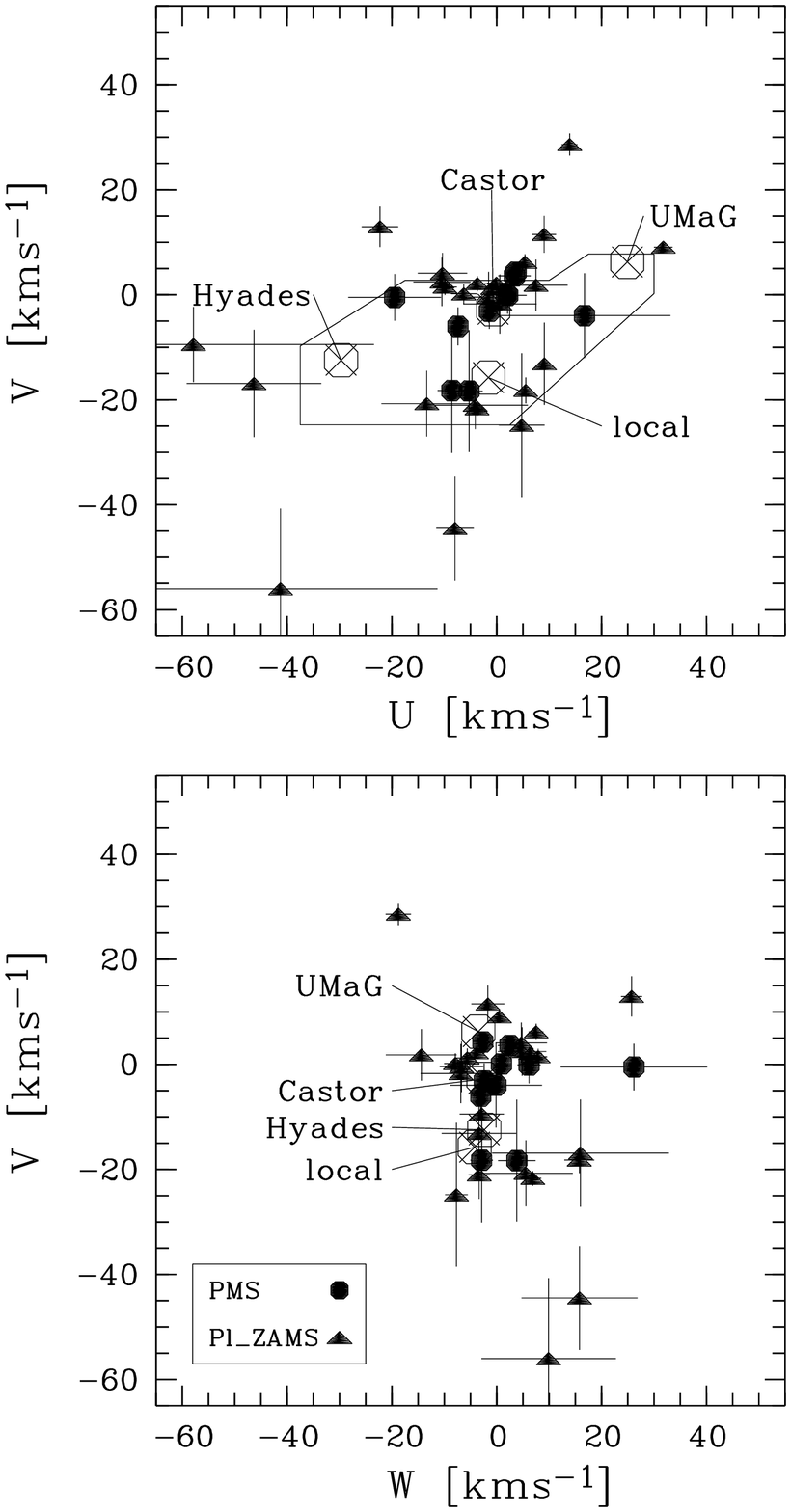}}
\caption[]{Upper panel: $U-V$ velocity diagram for the youngest age groups 
``PMS'' (circles) and  ``Pl\_ZAMS'' (triangles).
The solid line  encircles the region defined by Eggen (\cite{Eggen84}, \cite{Eggen89}) 
to contain the young disk population. Also
shown as large crossed circles are the $U$ and $V$ velocities of the 
Hyades supercluster, the Local Association (designated ``local''), the 
Castor MG, and the UMa MG. Lower panel: $W-V$ diagram for the same sample of stars. 
All velocities are in the LSR reference frame.
}
\label{UVeggen}
\end{figure}

\begin{table*} 
\tabcolsep=8pt
\centering
\caption[]{Mean (M-L) space velocity components, $\langle U\rangle, \langle V\rangle, 
\langle W\rangle$, and velocity dispersions, $\sigma_U$, $\sigma_V$, $\sigma_W$, 
of the different age groups in \kms .
} 
\begin{tabular}{lllllll} 
\noalign{\smallskip}     
\hline 
\hline 
\noalign{\smallskip}    
       & \multicolumn{1}{c}{$\langle U \rangle$} & \multicolumn{1}{c}{$\sigma_U$} &\multicolumn{1}{c}{$\langle V \rangle$} 
       &   \multicolumn{1}{c}{$\sigma_V$} & \multicolumn{1}{c}{$\langle W \rangle$} & \multicolumn{1}{c}{$\sigma_W$}\\ 
\noalign{\smallskip}    
\hline 
\noalign{\smallskip}    
PMS      & $-1.9 \pm 2.6$ & $ 4.2 \pm 3.1$ & $+0.2 \pm 2.2$ &$ 2.9 \pm 2.5$ & $+0.0 \pm 1.7$ & $2.4\pm 2.1$    \\
Pl\_ZAMS & $-1.5 \pm 4.1$ & $11.9 \pm 3.8$ & $-4.4 \pm 4.8$ &$14.6 \pm 4.6$ & $+1.1 \pm 3.2$ & $9.3 \pm 2.9$ \\
UMa      & $+18.2 \pm 6.6$& $12.1 \pm 11.1$ & $-1.2 \pm 10.8$&$22.1 \pm 10.6$& $-5.2 \pm 10.5$& $21.8 \pm  10.1$  \\
Hya+:& & & & &  &\\
\multicolumn{1}{r}{dwarfs} & $+5.5 \pm 7.0$ & $22.4 \pm 6.6$ & $-7.9 \pm 5.3$ &$16.3 \pm 4.6$ & $+6.4 \pm 8.8$ & $29.2 \pm 7.8$  \\
\multicolumn{1}{r}{giants} & $+4.5 \pm 11.7$ & $28.4 \pm 12.9$ & $-15.0 \pm 8.9$ &$19.2 \pm 8.1$ & $+0.3 \pm 5.8$ & $13.4 \pm 6.2$  \\
\noalign{\smallskip}    
\hline 
\noalign{\smallskip}    
\end{tabular}
\label{uvwmean}
\end{table*}

The space velocities components are plotted in Fig. \ref{UVW}. The plot contains 
stars of all age groups and also includes the evolved stars (giants).
Figure \ref{UVeggen} shows in an enlarged scale the  $V-U$, and $V-W$ diagrams 
for the two youngest stellar age groups only, i.e. PMS and Pl\_ZAMS stars.  

As can be seen in Fig. \ref{UVW} the filled symbols representing the youngest age
groups, PMS and Pl\_ZAMS, are more concentrated than the open symbols 
and the asterisks denoting the older age groups and giants, respectively. 
This can be tested by various statistical methods. 
First, we combined  on one hand 
the PMS and Pl\_ZAMS samples and  on the other hand the older stars and giants 
in order to create distributions of the space velocity $v_{\rm LSR} =
\sqrt{U^2+V^2+W^2}$ for the young and the old stars,
respectively.  
A one-dimensional two-sample Kolmogorov-Smirnov (K-S) test on these  
distributions yields a probability of $<1.6\,10^{-5}$ 
that they are drawn from the same parent distribution. 
Likewise, the K-S test on the PMS and the complementary non-PMS sample yields 
a probability of only $5\,10^{-4}$ for having the same distribution. 
Therefore, PMS and non-PMS stars also have different space velocity distributions. 
Contrary to this, with a probability of 0.31 PMS and Pl\_ZAMS stars have 
the same distribution.
An F-test on the individual velocity components $U$, $V$, and $W$ of 
the combined PMS-Pl\_ZAMS and the
older age groups shows that with a very low probability $P$ their 
distributions are drawn from the same parent distribution, namely 
$P_U = 0.006$, $P_V= 0.02$, and $P_W = 4\,10^{-6}$. In particular, the velocity
component perpendicular to the galactic plane, $W$, is significantly different
in the young and the old age groups (see below). 

In the following we will discuss mean velocities and velocity dispersions of the
different age groups. 
These  were calculated 
as maximum-likelihood (M-L) 
estimate 
which takes into account that the measurement errors are
different for each star. 
Following  Pryor \& Meylan (\cite{PryorMeylan93}) M-L estimates of the mean
velocity components $\langle v \rangle$ and dispersions $\sigma_v$ of $U, V$ and $W$ 
were obtained together with errors by assuming that the velocities are drawn from 
a normal distribution
\begin{equation}
 f(v_i) =  \frac{1}{\sqrt{2\,\pi (\sigma_{v}^2+
 \sigma_i^2})}\exp \left( -\frac{1}{2}\frac{(v_i - \langle v \rangle)^2}{(\sigma_v^2+ \sigma_i^2)}\right) 
\label{normal}
\end{equation}
with the individual velocity measurements $v_i$ and associated errors $\sigma_i$ of $U, V,$,
and, $W$, respectively. With the likelihood  function $\cal L$  defined as
\begin{equation}
{\cal L} =  \prod_{i=1}^{n}  f(v_i)
\label{likeli}
\end{equation}
the minimization of the test statistic $S = -2\,\ln {\cal L}$ then allows to derive the M-L estimates of 
$\langle v \rangle$ and  $\sigma_v$. Errors were 
calculated following  Pryor \& Meylan. 

In Table \ref{uvwmean} the mean space velocities and velocity dispersions 
of the different age groups 
are summarized. Clearly the PMS sample has the smallest velocity dispersions. 
For stars with weak or no lithium detection the dispersions are the largest. The  
``Hya+'' subsample  contains a significant fraction of older disk stars. This is 
particularly evident for the velocity component perpendicular to the 
galactic plane, $W$. 
Its dispersion increases from  $\sim2$\,\kms\ for the PMS sample to $\sim30$\,\kms\ 
for the old lithium weak sample. The increasing velocity dispersion with increasing
age reflects the effect of disk heating in the galaxy.

The PMS subsample in particular exhibits M-L mean space velocity components 
$(\langle U\rangle, \langle V \rangle, \langle W \rangle) = 
(-1.9\pm2.6, +0.2\pm2.2, +0.0\pm1.7)$\,\kms\ and
velocity dispersions of ($\sigma_U, \sigma_V, \sigma_W)  = (4.2\pm3.1, 2.9\pm2.5,  
2.4\pm2.1$)\,\kms . This suggests that the PMS stars are kinematically related and 
may even form a kinematical group, but of course, the sample is small and the 
indicated relation should be considered more as a working hypothesis to be tested with 
extended samples.  
At this point it should be noted that the 
PMS star B206 in area II interestingly has space velocity components similar 
to the stars in area I.
Unfortunately, no high resolution RV measurement is available for the second
PMS star in area II, the M4Ve dwarf B026. In order to obtain at least an estimate of its
space velocity components we measured the radial velocity using the low-resolution CAFOS
spectra and the emission lines of H$\alpha$, H$\beta$, H$\gamma$, and {\CaII}K. This
yielded $v_{\rm hel} = -12$\,\kms\  with an error of about 20\,\kms .
The resulting space velocity components are $U = +20\pm17$\,\kms , 
$V = +1\pm8$\,\kms , and $W = +2\pm9$\,\kms . 
Within the errors the velocities of B026 are consistent with the 
mean velocities of the PMS sample. But clearly, a more accurate RV measurement is needed 
for B026 to confirm that both Li-rich stars in area II belong 
to the same kinematical group as the corresponding stars in area I as indicated 
by the presently available data. 
Note also from Fig. \ref{EWLifield} that in areas I and II the numbers of Li-rich stars are higher 
than in the other areas.

Enlarged sections of the $U-V$ and $W-V$ dia\-gram are shown in Fig. 
\ref{UVeggen}  for the 35 stars of the two youngest age groups with measured 
space velocities. The $U-V$ dia\-gram in the upper panel
includes the limits of the region occupied by the young disk stars  as defined by Eggen 
(\cite{Eggen84}, \cite{Eggen89}). Indeed, as expected for a young stellar sample many, 
albeit not all, stars have $(U,V)$ velocities inside Eggen's box. 
Also indicated are the $(U,V)$ velocities of several 
young stellar kinematical groups: 
the Hyades supercluster,  the Ursa Major moving group (UMa MG), the Local 
Association (Pleiades MG), and the Castor moving group (Castor MG) 
(for references see e.g. Montes et al. \cite{Montesetal01}).

Figure \ref{UVeggen} suggests the existence of a kinematical subgroup in the 
combined PMS and Pl\_ZAMS sample, which contains 35 stars with measured 
$U$, $V$, and $W$ velocities. The group is concentrated  near 
the velocity of the Castor MG at the upper $V$ limit of Eggen's disk stars with 
17 of the 35 stars found within a radius of 10\,\kms\ around the velocity 
of the Castor MG.  
Six of these belong to the age group PMS and the rest to the Pl\_ZAMS group. 
The M-L mean velocities of the subgroup are 
$(U, V) = (-1.4\pm1.7, +2.3\pm0.9)$\,\kms , and the velocity
dispersions are $(\sigma_U, \sigma_V) = (4.3\pm1.0, 1.2\pm0.6)$\,\kms . 
In $V$ the group of 17 stars is 
somewhat off the Castor MG for which Palou\v{s} \& Piskunov (\cite{PalousPiskunov85}) 
give ($U,V) = (-0.7\pm3.5, -2.8\pm2.4)$\,\kms . 
Given the relatively small number of data points we may ask whether this
concentration is due to a chance coincidence in an actually random distribution. 
We tested this possibility for the null 
hypothesis that the true underlying distribution of
velocities in the $U-V$ plane is random within a given circle around 
the origin. In a Monte Carlo simulation we calculated a large number of random
velocity vectors in the $U-V$ plane and counted the number of cases in which 
we found 17 stars within 
10\,\kms\ around the Castor MG velocity. For a random velocity distribution within
a radius of $35$\,\kms\ containing 90\% of the 35 stars, i.e. 31 stars, 
these simulations showed that we can reject the null hypothesis on a 
high significance level of $>99.8$\%. The test radius of $35$\,\kms\ may be 
too small because it excludes 10\% of the stars. Increasing the radius 
leads however to even higher significance levels. Decreasing the radius only leads to
significance levels of $<99$\% if the random distribution is calculated 
for radii smaller than $\sim$20\,\kms\ which contains $\le$65\% of the  
PMS-Pl\_ZAMS stars. Therefore we are lead to the conclusion that with a 
very high probability the concentration of velocities vectors in the $U-V$ plane is 
not a chance coincidence.  

An interesting feature is the accumulation of the 6 
``PMS'' stars around a mean velocity of ($U,V) = 
(+0.0\pm2.9, +1.2\pm2.8)$\,\kms . This is not far 
from the velocity of the Castor  MG (see above), 
but clearly distinct from the Local Association which has 
($U,V) = (-1.6, -15.8)$\,\kms\ (Montes et al. \cite{Montesetal01}). 
The velocity dispersions of this subgroup of PMS stars are 
$(\sigma_U, \sigma_V) = (3.5\pm1.7, 2.5\pm2.1)$\,\kms .
Two of the remaining PMS stars are found near
the velocity of the Local Association together with a loose accumulation 
of some 5 or 6 further stars from the Pl\_ZAMS age group. A relation of these stars 
with the Local Association may exist,
but the errors and the scatter of the velocity vectors are quite large.

The $W-V$ diagram displayed in the lower panel of Fig. \ref{UVeggen} shows a 
similar trend in the distribution of the velocity vectors as in the $U-V$ diagram,  
that is most PMS stars and many Pl\_ZAMS stars are kinematically distinct from 
the Local Association. 

\section{Conclusions}
\label{sum}
We have investigated the characteristics of an X-ray selected sample from the RASS of 
high-galactic latitude field stars
comprising 56 G, 86 K, and  56 M type stars. Spectroscopic low/medium 
and high resolution follow-up observations were obtained for 
95\% of the G-K stars and for 77\% of the M stars. 

Spectroscopic luminosity classification of the G-K stars 
based on the high resolution spectroscopy showed that 88\% of the G-K stars are 
main-sequence  stars or subgiants of luminosity classes V and IV, respectively. 
From IR photometric classification we concluded that all M stars are dwarf stars.

Significant lithium absorption lines were detected in a large fraction of
stars with equivalent widths and abundances, respectively, above the level of the 
Hyades in about 50\% of the stars. 
For the age distribution of the high-galactic latitude coronal sample this means
that about half of the G-K stars are younger than the Hyades.
About 25\% of the G-K stars have an age comparable to that of the Pleiades, i.e.
$\sim100$\,Myr. A small fraction of less than 10\% of the G-K stars is younger 
than the Pleiades. Most PMS stars, i.e. 8 out of 10, are located 
in area~I. Only two PMS stars are found in area II and none in the remaining
areas. This suggests a possible relation of the high-$|b|$  PMS stars 
to the Gould Belt indicated in  Fig. \ref{area_gal}. However, the subsample formed 
by combining the stellar age groups PMS and Pl\_ZAMS is  
spatially distributed in all directions covered by our study areas. At the same time 
half 
of its members show similar kinematical parameters independent of 
spatial location. This questions the relation to the Gould Belt. Rather, the space 
velocities suggest that these stars are members of a loose moving group with a mean 
velocity close to that of the Castor MG.  
For the Castor MG an age of 200$\pm$100\,Myr has been derived by
Barrado y Navascu\'es (\cite{Barrado98}). This would still be consistent with the Pl\_ZAMS 
group. If some of the PMS stars are indeed kinematically related to the Castor MG
this would indicate a large age spread in this moving group as they appear 
to be younger than 100\,Myr, maybe even as young as $\sim30$\,Myr.

\acknowledgements{We would like to thank the Deutsche For\-schungs\-gemeinschaft 
for granting travel funds (Zi\,420/3-1, 5-1, 6-1, 7-1). We further thank the staff at the German-Spanish 
Astronomical Centre, Calar Alto, in particular Santos Petraz, for carrying out part of the 
observing programme in service mode. This publication makes use of data products from the 
Two Micron All Sky Survey, which is a joint project of the University of Massachusetts and 
the Infrared Processing and Analysis Center/California Institute of Technology, funded by the 
National Aeronautics and Space Administration and the National Science Foundation. This research 
has made use of the SIMBAD and VIZIER databases, operated at CDS, Strasbourg, France.
}

\appendix
\section{Parameters of the sample and results}
\label{results}

\begin{figure*}[h]
\resizebox{\hsize}{!}{\includegraphics[angle=0,bbllx=75pt,bblly=120pt,bburx=580pt,bbury=610pt,clip=]{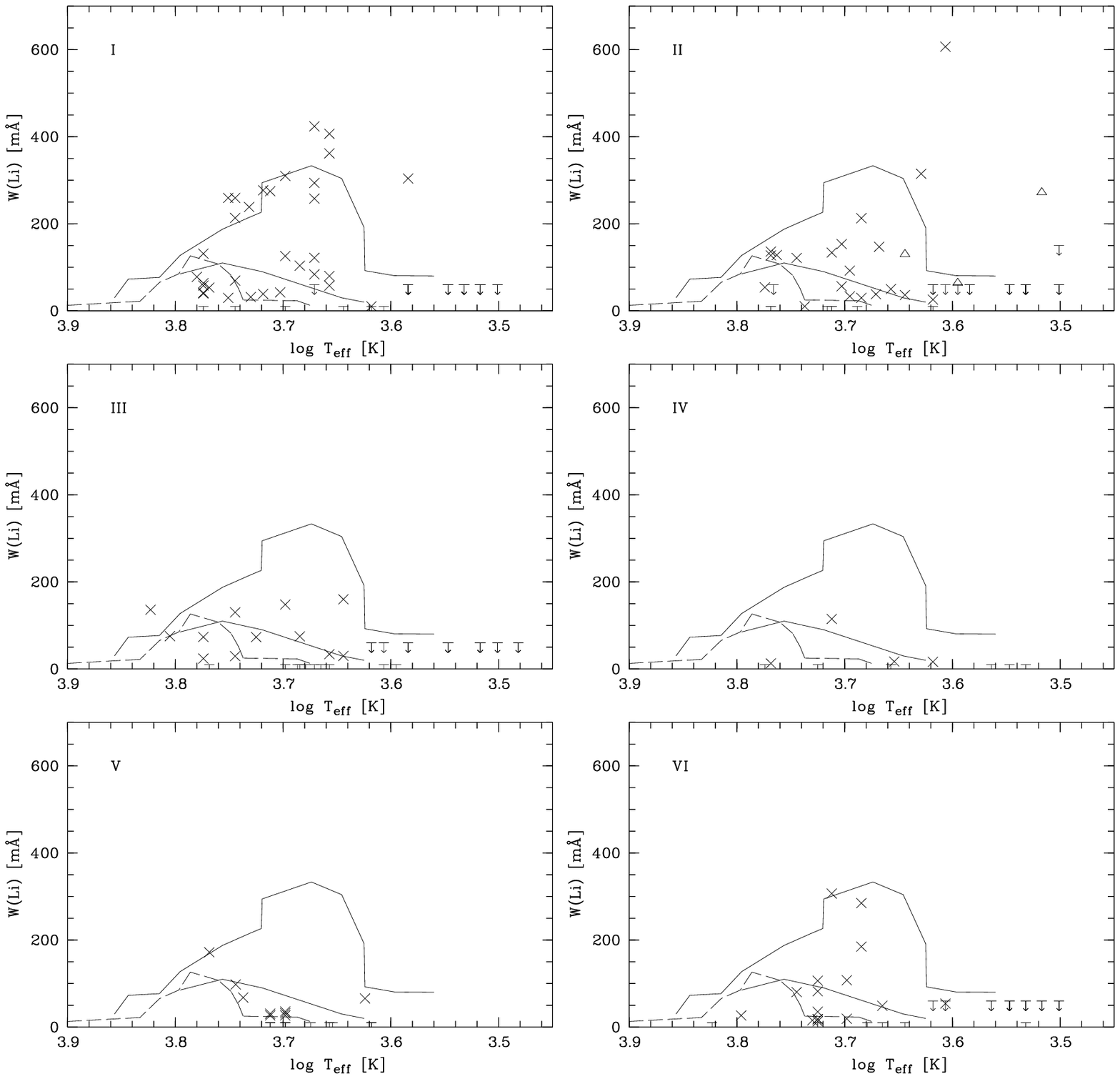}}
\caption[]{Equivalent widths of \LiI$\lambda6708$ as a function of $\teff$ plotted for each study
area individually. The solid and dashed lines have the same meaning as in Fig. \ref{soder}.
}
\label{EWLifield}
\end{figure*}

Figure \ref{EWLifield} displays the measured lithium equivalent widths for each study area 
separately.

Table \ref{vtangent} summarizes  for the individual study areas the mean tangential velocities 
in galactic coordinates, $v_l$ and $v_b$, and their dispersions as calculated in Sect. \ref{propermotions}. 
The mean solar velocity components, $\langle v_{l,\sun} \rangle$ and $\langle v_{b,\sun} \rangle$, 
are also given for each study area.
Furthermore, for each age group  mean distances $\langle d \rangle$ are given.

\begin{table*} 
\centering
\caption[]{Mean tangential velocities and dispersions (in \kms ) in galactic coordinates 
for stars of age groups  PMS, Pl\_ZAMS, UMa, Hya+ (split into dwarfs and giants),
and M stars without lithium detection for the individual study areas. 
The first two lines give the  average solar velocity for each study area.
The mean distance $\langle d \rangle$ and its scatter 
are given in pc. The number of stars used for the calculation of the mean values 
are given in parentheses. 
} 
\begin{tabular}{lllllll} 
\noalign{\smallskip}     
\hline 
\hline 
\noalign{\smallskip}    
   area               & \multicolumn{1}{c}{I}                  & \multicolumn{1}{c}{II}               & \multicolumn{1}{c}{III}               &   \multicolumn{1}{c}{IV}               & \multicolumn{1}{c}{V}                  & \multicolumn{1}{c}{VI} \\
\noalign{\smallskip}    
\hline 
\noalign{\smallskip}    
$\langle v_{l,\sun} \rangle$& \multicolumn{1}{c}{$-4$}&\multicolumn{1}{c}{$-9$}&\multicolumn{1}{c}{$+8$}  &\multicolumn{1}{c}{$-5$} &\multicolumn{1}{c}{$-11$} &\multicolumn{1}{c}{$-8$}     \\
$\langle v_{b,\sun} \rangle$& \multicolumn{1}{c}{$-0$}&\multicolumn{1}{c}{$+10$}&\multicolumn{1}{c}{$+11$} &\multicolumn{1}{c}{$+10$}&\multicolumn{1}{c}{$+4$}  &\multicolumn{1}{c}{$+11$}     \\
PMS: & & & & & &   \\
\multicolumn{1}{r}{$\langle v_l \rangle$} & $+9 \pm 9 (8)$    & $+7 \pm 2$ (2)    &\multicolumn{1}{c}{--}                &\multicolumn{1}{c}{--} & \multicolumn{1}{c}{--}                 & \multicolumn{1}{c}{--}\\
\multicolumn{1}{r}{$\langle v_b \rangle$} & $+5 \pm 11$ (8)   & $-3 \pm 1$ (2)    &\multicolumn{1}{c}{--}                &\multicolumn{1}{c}{--} & \multicolumn{1}{c}{--}                 & \multicolumn{1}{c}{--}\\
\multicolumn{1}{r}{$\langle d \rangle$}   & $121 \pm 53$ (8)  & $51 \pm 20$ (2)   &\multicolumn{1}{c}{--}                &\multicolumn{1}{c}{--} & \multicolumn{1}{c}{--}                 & \multicolumn{1}{c}{--} \\
Pl\_ZAMS: & & & & & \\
\multicolumn{1}{r}{$\langle v_l \rangle$} & $+11 \pm 13$ (13) & $+13 \pm 4$ (8)   & $-6 \pm 8$ (6)    &\multicolumn{1}{c}{--}                &$+52 \pm 26$ (3)    & $+2 \pm 4$ (6)\\
\multicolumn{1}{r}{$\langle v_b \rangle$} & $-1 \pm 15$ (13)  & $-5 \pm 7$  (8)   & $-19 \pm 11$ (6)  &\multicolumn{1}{c}{--}                &$+7 \pm 19$ (3)     & $-6 \pm 12$ (6)\\
\multicolumn{1}{r}{$\langle d \rangle$}   & $118 \pm 56$ (14) & $117 \pm 50$ (9)  & $91 \pm 39$ (7)   &\multicolumn{1}{c}{--}                & $226 \pm 191$ (3)  & $101 \pm 77$ (6)\\
UMa: & & & & &  \\
\multicolumn{1}{r}{$\langle v_l \rangle$} & $+4 \pm 13$ (2)   & \multicolumn{1}{c}{--}                &\multicolumn{1}{c}{--}                &                & $-24 \pm 42$ (6)   & $-10 \pm 18$ (6)\\
\multicolumn{1}{r}{$\langle v_b \rangle$} & $-10 \pm 12$ (2)  & \multicolumn{1}{c}{--}                &\multicolumn{1}{c}{--}                &                & $-10 \pm 21$ (6)    & $-7 \pm 8$ (6)\\
\multicolumn{1}{r}{$\langle d \rangle$}   & $323 \pm 191$ (3) & \multicolumn{1}{c}{--}                &\multicolumn{1}{c}{--}               &                &  $332 \pm 252$ (6) &$240 \pm 184$ (6) \\
Hya+: & & & & &  \\
\multicolumn{1}{r}{dwarfs:} & & & & &  \\
\multicolumn{1}{r}{$\langle v_l \rangle$} & $+3 \pm 12$ (8) & $+4 \pm 17$ (6)  & $-6 \pm 39$ (9)  & $+11 \pm 23$ (4)   &$+12 \pm 15$ (3)    & $-11 \pm 45$ (4)\\
\multicolumn{1}{r}{$\langle v_b \rangle$} & $-5 \pm 15$ (8)  & $-9 \pm 9$ (6)  &$-15\pm 16$ (9)  & $-2 \pm 26$ (4)  &$-3 \pm 8$ (3)    & $-24 \pm 36$ (4)\\
\multicolumn{1}{r}{$\langle d \rangle$}   & $82 \pm 46$ (8)  & $132 \pm 165$ (6)& $90 \pm 56$ (9)  & $204 \pm 102$ (4) & $56 \pm 45$ (3) &$201 \pm 245$ (4) \\
\multicolumn{1}{r}{giants:} & & & & &  \\
\multicolumn{1}{r}{$\langle v_l \rangle$}  & $-2 \pm 12$ (2)   & $+11 \pm 93$ (10) &\multicolumn{1}{c}{--} & \multicolumn{1}{c}{--}                &$+6 \pm 21$ (2)    &\multicolumn{1}{c}{--}\\
\multicolumn{1}{r}{$\langle v_b \rangle$}  & $+7 \pm 7$ (2)    & $-8 \pm 42$ (10)  &\multicolumn{1}{c}{--} & \multicolumn{1}{c}{--}                &$+15 \pm 19$ (2)   &\multicolumn{1}{c}{--}\\
\multicolumn{1}{r}{$\langle d \rangle$ }   & $124 \pm 26$ (2)  & $601 \pm 810$ (10)&\multicolumn{1}{c}{--}& \multicolumn{1}{c}{--}                 & $175 \pm 101$ (2) &\multicolumn{1}{c}{--} \\
M stars: & & & & &  \\
\multicolumn{1}{r}{$\langle v_l \rangle$} & $+12 \pm 10$ (12) & $0 \pm 21$ (5)    & $-8 \pm 16$ (10)  & $-3 \pm 27$ (6)    & $+6 \pm 27$ (8)    & $+10 \pm 18$ (7)\\
\multicolumn{1}{r}{$\langle v_b \rangle$} & $+0 \pm 9$  (12)  & $+5 \pm 16$ (5)   & $-15 \pm 15$ (10) & $-44 \pm 35$ (6)   &$-10 \pm 27$ (8)    & $-19 \pm 11$ (7)\\
\multicolumn{1}{r}{$\langle d \rangle$}   & $43 \pm 30$ (14)  & $60 \pm 24$ (6)   & $56 \pm 32$ (11)  & $53 \pm 27$ (6)    & $59 \pm 23$ (8)    & $43 \pm 15$ (7)\\
\noalign{\smallskip}    
\hline 
\noalign{\smallskip}    
\end{tabular}
\label{vtangent}
\end{table*}

The following Tables \ref{basic} to \ref{resulttab3} are available as Online Material only.

In Table \ref{basic} the basic parameters of the stellar sample are listed. 
The field and RASS names were taken from Paper III. 
Coordinates R.A.\,(2000) and DEC.\,(2000) of the optical counterparts are in succession either from 
Tycho-2, GSC-I, or GSC-II, whatever the source  is for the $V$ magnitude listed in column ``V''.
In column ``Sp.type'' the revised spectral types with luminosity
class are given for objects with new high resolution observations. 
Otherwise, spectral types from Paper III are given. 
Spectroscopic binaries are flagged by ``SB2''. B160 is a triple system (SB3).
The flux ratio $\log f_{\rm x}/f_V$ is given
for the revised V magnitudes and the RASS fluxes from Paper III. 
X-ray luminosities $L_{\rm x}$ were calculated 
using the distances listed in column ``dist''. The distances are  flagged by ``S'', ``H'', ``T''
or ``I'', depending on whether they were derived  from spectroscopy,  Hipparcos, 
trigonometric parallaxes, or from  the infrared colours, respectively. Distance estimates 
obtained by assuming luminosity class V are flagged by ``M''. They should be considered 
as lower limits only.

Table \ref{resulttab2} lists the kinematical parameters. Heliocentric radial velocities 
and errors for single stars and for the primary component of spectroscopic binaries
are given in columns $v_{\rm hel,1}$ and $\sigma_1$, respectively. For binaries columns
$v_{\rm hel,2}$  and $\sigma_2$ contain the heliocentric radial velocity and error 
of the secondary component, respectively. Proper motions and associated errors are
listed in columns $\mu_{\alpha}\,\cos\delta$ and $\sigma_{\mu_{\alpha}}$ for 
right ascension, and  $\mu_{\delta}$ and $\sigma_{\mu_{\delta}}$ for declination, 
respectively.
The source catalog of the proper motions is denoted by
respective  flags: TY = Tycho-2, HI = Hipparcos, UC = UCAC2, US = USNO-B1.0, PP = PPM, ST =
STARNET, TR = TRC, NL = NLPM1, CA = Carlsberg Meridian Catalogs.  Also given are the 
galactic velocity components $U$, $V$, and $W$ in the LSR frame 
with errors $\sigma_U, \sigma_V,$ and $\sigma_W$, respectively.  
If the errors were larger than 30\,\kms\ the space velocity components were omitted. 

Table \ref{resulttab3} lists lithium data and rotational velocities.  
Equivalent widths of \LiI\,$\lambda6708$ are listed in column $W($\LiI ). 
Flags ``h'', ``l''or ``m'' denote high-, low- or medium resolution measurements,
respectively. Lithium abundances derived from $W($\LiI ) for the effective temperatures given in
column $\teff $ are listed in column $\log N$(Li). 
The last two columns list the rotational velocities, $v_{\rm rot_{1}}$ for single stars or primary
components of binaries, and in the latter case $v_{\rm rot_{2}}$ for the secondary component.

\Online
\begin{table*}
\centering
\caption[]{Basic optical and X-ray parameters of the sample of G-, K-, and M-type stars. 
} 
\label{basic}
\tabcolsep=5pt
\centering
\begin{tabular}{lllllllllllrl} 
\noalign{\smallskip}     
\hline 
\hline 
\noalign{\smallskip}    
field & RASS name        &R.A.   & DEC.  &Sp. type&  & $V$ & $\log f_{\rm x}/f_V$
&dist. &  & $\log L_{\rm X}$ \\
      &              &(2000) & (2000) &     &          &     &                   
      &[pc]  &  &[erg\,s$^{-1}$]  \\
\noalign{\smallskip}    
\hline 
\noalign{\smallskip}    
A001    &RXJ0328.2$+$0409&03:28:14.9&$+$04:09:48&K0      &SB2&    9.63&$ -2.08$ &  76:&M&    30.23                 \\ 
A007    &RXJ0330.7$+$0305&03:30:43.5&$+$03:05:48&K1      &SB2&   10.80&$ -2.62$ & 119:&M&    29.62                 \\ 
A010    &RXJ0331.1$+$0713&03:31:08.4&$+$07:13:25&K4Ve    &   &   10.86&$ -2.07$ &  62 &S&    29.88                 \\ 
A013    &RXJ0331.4$+$0455&03:31:25.7&$+$04:55:08&M4e     &   &   13.50&$ -1.53$ &  28 &I&    28.67                 \\ 
A028    &RXJ0336.5$+$0726&03:36:34.3&$+$07:26:20&G9V     &   &   10.89&$ -2.56$ & 105 &S&    29.83                 \\ 
A030    &RXJ0336.6$+$0329&03:36:40.7&$+$03:29:22&M5Ve    &   &   13.86&$ -1.62$ &  14 &T&    27.86                 \\
\end{tabular}
\end{table*}

\addtocounter{table}{-1}
\begin{table*}
\tabcolsep=5pt
\centering
\caption[]{continued.
} 
\begin{tabular}{lllllllllll} 
\noalign{\smallskip}     
\hline 
\hline 
\noalign{\smallskip}    
field & RASS name        &R.A.   & DEC.  &Sp. type&  & $V$ & $\log f_{\rm x}/f_V$
&dist. &  & $\log L_{\rm X}$ \\
      &              &(2000) & (2000) &     &          &     &                   
      &[pc]  &  &[erg\,s$^{-1}$]   \\
\noalign{\smallskip}    
\hline 
\noalign{\smallskip}    
B004    &RXJ0642.7$+$6405&06:42:46.2&$+$64:05:45&G5II    &   &    7.61&$ -4.17$ & 955 &S&    31.46                 \\ 
B008    &RXJ0648.5$+$6639&06:48:35.9&$+$66:39:14&G5      &SB2&   10.88&$ -2.86$ & 198:&M&    29.78                 \\ 
B013    &RXJ0701.0$+$6541&07:01:02.1&$+$65:41:47&K2Ve    &   &   12.12&$ -2.29$ & 140 &S&    29.86                 \\ 
B018    &RXJ0704.0$+$6214&07:04:05.8&$+$62:15:01&K5Ve    &   &   12.27&$ -2.35$ & 105 &S&    29.49                 \\ 
B018    &RXJ0704.0$+$6214&07:04:07.5&$+$62:14:42&K9      &   &   14.68&$ -1.38$ & 164:&M&    29.88                 \\ 
B023    &RXJ0707.0$+$5752&07:07:01.8&$+$57:52:35&K7e     &   &   12.00&$ -2.17$ &  68:&M&    29.41                 \\
\end{tabular}
\end{table*}

\addtocounter{table}{-1}
\begin{table*}
\tabcolsep=5pt
\centering
\caption[]{continued.
} 
\begin{tabular}{lllllllllll} 
\noalign{\smallskip}     
\hline 
\hline 
\noalign{\smallskip}    
field & RASS name        &R.A.   & DEC.  &Sp. type&  & $V$ & $\log f_{\rm x}/f_V$
&dist. &  & $\log L_{\rm X}$ \\
      &              &(2000) & (2000) &     &          &     &                   
      &[pc]  &  &[erg\,s$^{-1}$]   \\
\noalign{\smallskip}    
\hline 
\noalign{\smallskip}    
C095    &RXJ1041.3$-$0144&10:41:24.3&$-$01:44:28&K1IV    &   &    6.25&$ -4.17$ &  34 &H&    29.11                 \\ 
C106    &RXJ1041.9$+$0208&10:41:58.9&$+$02:08:43&M2e     &   &   15.05&$ -1.33$ &  99 &I&    29.35                 \\ 
C120    &RXJ1045.0$+$0043&10:45:05.0&$+$00:43:33&M5e     &   &   15.23&$ -1.28$ &  46 &I&    28.66                 \\ 
C125    &RXJ1047.8$-$0113&10:47:51.7&$-$01:13:31&K0IV    &   &    9.87&$ -2.68$ & 226 &S&    30.79                 \\ 
C143    &RXJ1051.3$-$0734&10:51:24.1&$-$07:34:02&K2V     &   &   10.59&$ -2.94$ &  69 &S&    29.21                 \\ 
C146    &RXJ1051.8$+$0235&10:51:56.3&$+$02:35:54&K8e     &   &   14.03&$ -1.68$ & 147:&M&    29.75                 \\
\end{tabular}
\end{table*}

\addtocounter{table}{-1}
\begin{table*}
\tabcolsep=5pt
\centering
\caption[]{continued.
} 
\begin{tabular}{lllllllllll} 
\noalign{\smallskip}     
\hline 
\hline 
\noalign{\smallskip}    
field & RASS name        &R.A.   & DEC.  &Sp. type&  & $V$ & $\log f_{\rm x}/f_V$
&dist. &  & $\log L_{\rm X}$ \\
      &              &(2000) & (2000) &     &          &     &                   
      &[pc]  &  &[erg\,s$^{-1}$]   \\
\noalign{\smallskip}    
\hline 
\noalign{\smallskip}    
E170    &RXJ1726.4$+$7422&17:26:29.2&$+$74:21:42&M5e     &   &   14.61&$ -1.91$ &  34 &I&    28.03                 \\ 
E179    &RXJ1728.1$+$7239&17:28:12.4&$+$72:39:23&K4IVe   &   &   11.38&$ -2.48$ & 445 &S&    30.97                 \\ 
E221    &RXJ1732.6$+$7413&17:32:41.3&$+$74:13:38&G9III   &   &    6.63&$ -3.11$ & 103 &H&    30.98                 \\ 
E256    &RXJ1736.2$+$7152&17:36:13.2&$+$71:52:42&K4V     &   &    8.55&$ -4.18$ &  19 &H&    27.66                 \\ 
E262    &RXJ1736.9$+$7420&17:36:54.7&$+$74:20:25&K0V-IV  &   &   10.10&$ -3.35$ & 130 &S&    29.55                 \\ 
F002    &RXJ2152.0$+$1436&21:52:01.9&$+$14:36:06&K1IV    &   &    8.30&$ -3.98$ & 110 &S&    29.49                 \\
\end{tabular}
\end{table*}

\begin{table*}
\centering
\caption[]{Kinematical parameters. 
} 
\label{resulttab2}
\tabcolsep=5pt
\centering
\begin{tabular}{llllllllllllll} 
\noalign{\smallskip}     
\hline 
\hline 
\noalign{\smallskip}    
field & Sp. type        &$v_{\rm hel,1}$&$\sigma{_1}$ & $v_{\rm hel,2}$&$\sigma{_2}$ &
$\mu_{\alpha}\cos\delta$         & $\sigma_{\mu_{\alpha}}$ &
$\mu_{\delta}$ & $\sigma_{\mu_{\delta}}$& &
$U\pm\sigma_U$  & $V\pm\sigma_V$ & $W\pm\sigma_W$ 
\\
      &                  &   \multicolumn{2}{c}{[\kms ] }     &  \multicolumn{2}{c}{[\kms ] }          &
\multicolumn{2}{c}{[mas\,yr$^{-1}$]}  & 
\multicolumn{2}{c}{[mas\,yr$^{-1}$]}  & & 
\multicolumn{3}{c}{[\kms ]}      
\\
\noalign{\smallskip}    
\hline 
\noalign{\smallskip}    
A001    &K0              &$   +12         $&     2 &$   +13$&     2&  +46.6&    1.1&  -35.2&    1.1&UC&$          * $&$          * $&$          * $\\ 
A007    &K1              &$   -56         $&     2 &$  +115$&     2&  +44.2&    1.5&  -84.6&    2.0&UC&$          * $&$          * $&$          * $\\ 
A010    &K4Ve            &$   +10         $&     1 &$     *$&     *&  +13.7&    1.5&  -13.1&    1.3&UC&$  +1.8 \pm   1.6 $&$  +0.1 \pm   2.8 $&$  +0.9 \pm   1.3 $\\ 
A013    &M4e             &$     *         $&     * &$     *$&     *&  +54.0&    2.0&  +20.0&    3.0&US&$          * $&$          * $&$          * $\\ 
A028    &G9V             &$    +0         $&     3 &$     *$&     *&  -35.1&    2.9&  -17.7&    2.5&UC&$ +22.0 \pm   6.2 $&$ +10.2 \pm   2.8 $&$  -7.8 \pm   7.7 $\\ 
A030    &M5Ve            &$     *         $&     * &$     *$&     *& +114.0&    2.0& -124.0&    2.0&US&$          * $&$          * $&$          * $\\
\end{tabular}
\end{table*}

\addtocounter{table}{-1}
\begin{table*}
\tabcolsep=5pt
\centering
\caption[]{continued.
} 
\begin{tabular}{llllllllllllll} 
\noalign{\smallskip}     
\hline 
\hline 
\noalign{\smallskip}    
field & Sp. type        &$v_{\rm hel,1}$&$\sigma{_1}$ & $v_{\rm hel,2}$&$\sigma{_2}$ &
$\mu_{\alpha}\cos\delta$         & $\sigma_{\mu_{\alpha}}$ &
$\mu_{\delta}$ & $\sigma_{\mu_{\delta}}$&&
$U\pm\sigma_U$  & $V\pm\sigma_V$ & $W\pm\sigma_W$ 
\\
      &                  &   \multicolumn{2}{c}{[\kms ] }     &  \multicolumn{2}{c}{[\kms ] }          &
\multicolumn{2}{c}{[mas\,yr$^{-1}$]}  & 
\multicolumn{2}{c}{[mas\,yr$^{-1}$]}  & &
\multicolumn{3}{c}{[\kms ]}      
 \\
\noalign{\smallskip}    
\noalign{\smallskip}    
\hline 
\noalign{\smallskip}    
B004    &G5II            &$   +10         $&     3 &$     *$&     *&   -5.7&    1.3&  -36.5&    1.4&TY&$ -94.3 \pm  48.7 $&$-112.4 \pm  60.8 $&$ -49.8 \pm  31.7 $\\ 
B008    &G5              &$   -61         $&     3 &$   +27$&     2&   +5.8&    1.8&  -12.6&    1.9&TY&$          * $&$          * $&$          * $\\ 
B013    &K2Ve            &$   +67         $&     3 &$     *$&     *&  -18.0&    4.0& -140.0&    2.0&US&$ -97.1 \pm  27.7 $&$ -35.9 \pm  35.6 $&$ +11.3 \pm  13.2 $\\ 
B018    &K5Ve            &$   +15         $&     3 &$     *$&     *&  -16.0&    3.0&  -24.0&    2.0&US&$ -10.4 \pm   4.7 $&$  +4.1 \pm   3.9 $&$  +4.7 \pm   4.9 $\\ 
B018    &K9              &$     *         $&     * &$     *$&     *&  -18.0&    2.0&  -20.0&    4.0&US&$          * $&$          * $&$          * $\\ 
B023    &K7e             &$     *         $&     * &$     *$&     *&  -42.0&    1.0&  -42.0&    3.0&US&$          * $&$          * $&$          * $\\
\end{tabular}
\end{table*}

\addtocounter{table}{-1}
\begin{table*}
\tabcolsep=5pt
\centering
\caption[]{continued.
} 
\begin{tabular}{llllllllllllll} 
\noalign{\smallskip}     
\hline 
\hline 
\noalign{\smallskip}    
field & Sp. type        &$v_{\rm hel,1}$&$\sigma{_1}$ & $v_{\rm hel,2}$&$\sigma{_2}$ &
$\mu_{\alpha}\cos\delta$         & $\sigma_{\mu_{\alpha}}$ &
$\mu_{\delta}$ & $\sigma_{\mu_{\delta}}$&&
$U\pm\sigma_U$  & $V\pm\sigma_V$ & $W\pm\sigma_W$ 
\\
      &                  &   \multicolumn{2}{c}{[\kms ] }     &  \multicolumn{2}{c}{[\kms ] }          &
\multicolumn{2}{c}{[mas\,yr$^{-1}$]}  & 
\multicolumn{2}{c}{[mas\,yr$^{-1}$]}  & &
\multicolumn{3}{c}{[\kms ]}      
 \\
\noalign{\smallskip}    
\noalign{\smallskip}    
\hline 
\noalign{\smallskip}    
C095    &K1IV            &$   +42         $&     1 &$     *$&     *& -137.3&    0.4& -119.2&    0.4&TY&$  -8.3 \pm   0.5 $&$ -41.1 \pm   1.4 $&$ +17.9 \pm   1.6 $\\ 
C106    &M2e             &$     *         $&     * &$     *$&     *&  -66.0&    4.0&  -14.0&    1.0&US&$          * $&$          * $&$          * $\\ 
C120    &M5e             &$     *         $&     * &$     *$&     *&  -34.0&    6.0&  +34.0&    3.0&US&$          * $&$          * $&$          * $\\ 
C125    &K0IV            &$    +7         $&     1 &$     *$&     *&  +20.0&    1.1&  -57.5&    1.5&UC&$ +57.1 \pm  24.1 $&$ -38.6 \pm  19.8 $&$  -6.6 \pm   9.8 $\\ 
C143    &K2V             &$    +4         $&     1 &$     *$&     *&  -40.5&    1.2&  +14.4&    1.2&UC&$  -3.7 \pm   6.9 $&$  +2.1 \pm   1.5 $&$  +6.4 \pm   2.2 $\\ 
C146    &K8e             &$     *         $&     * &$     *$&     *&  -54.0&    1.0&   -6.0&    4.0&US&$          * $&$          * $&$          * $\\
\end{tabular}
\end{table*}

\addtocounter{table}{-1}
\begin{table*}
\tabcolsep=5pt
\centering
\caption[]{continued.
} 
\begin{tabular}{llllllllllllll} 
\noalign{\smallskip}     
\hline 
\hline 
\noalign{\smallskip}    
field & Sp. type        &$v_{\rm hel,1}$&$\sigma{_1}$ & $v_{\rm hel,2}$&$\sigma{_2}$ &
$\mu_{\alpha}\cos\delta$         & $\sigma_{\mu_{\alpha}}$ &
$\mu_{\delta}$ & $\sigma_{\mu_{\delta}}$&&
$U\pm\sigma_U$  & $V\pm\sigma_V$ & $W\pm\sigma_W$ 
\\
      &                  &   \multicolumn{2}{c}{[\kms ] }     &  \multicolumn{2}{c}{[\kms ] }          &
\multicolumn{2}{c}{[mas\,yr$^{-1}$]}  & 
\multicolumn{2}{c}{[mas\,yr$^{-1}$]}  & &
\multicolumn{3}{c}{[\kms ]}      
 \\
\noalign{\smallskip}    
\noalign{\smallskip}    
\hline 
\noalign{\smallskip}    
E170    &M5e             &$     *         $&     * &$     *$&     *&  +14.0&    1.0&   -6.0&    4.0&US&$          * $&$          * $&$          * $\\ 
E179    &K4IVe           &$   -39         $&     1 &$     *$&     *&  -18.7&    2.4&  +27.6&    2.5&TY&$ -41.3 \pm  29.9 $&$ -56.0 \pm  15.3 $&$  +9.9 \pm  12.8 $\\ 
E221    &G9III           &$   -16         $&     1 &$     *$&     *&  -65.0&    1.0&  +30.4&    1.2&TY&$  -1.9 \pm   1.5 $&$ -27.5 \pm   2.3 $&$ +23.2 \pm   2.3 $\\ 
E256    &K4V             &$   -26         $&     1 &$     *$&     *&  +90.9&    1.2&  -36.8&    1.2&TY&$ +18.9 \pm   0.4 $&$ -11.5 \pm   1.7 $&$ -12.9 \pm   1.1 $\\ 
E262    &K0V-IV          &$    +9         $&     1 &$     *$&     *&  +28.1&    1.8&   +8.9&    1.9&TY&$  +3.7 \pm   2.8 $&$ +20.9 \pm   4.5 $&$  -3.6 \pm   7.8 $\\ 
F002    &K1IV            &$   +13         $&     1 &$     *$&     *&   -9.9&    0.8&  -41.3&    0.7&UC&$ +30.3 \pm   8.1 $&$  +4.3 \pm   5.9 $&$  -8.8 \pm   5.0 $\\
\end{tabular}
\end{table*}

\begin{table*}
\centering
\caption[]{Lithium abundances and rotational velocities. 
} 
\label{resulttab3}
\tabcolsep=5pt
\centering
\begin{tabular}{llllllrl} 
\noalign{\smallskip}     
\hline 
\hline 
\noalign{\smallskip}    
field & Sp. type  &$T_{\rm eff}$  &$W$(\LiI ) &  & 
$\log N$(Li) 
&\multicolumn{1}{l}{$v_{\rm rot,1}$}&$v_{\rm rot,2}$\\
      &           &  [K]         &  [m\AA ]  &  &                & [\kms ] &       [\kms ] \\     
\noalign{\smallskip}    
\hline 
\noalign{\smallskip}    
A001    &K0              &$  5152$&   $   $ 275 & h & $     3.12   $ &$   96$&$    7$ \\ 
A007    &K1              &$  4989$&   $ < $  10 & h & $ <   0.68   $ &$   30$&$   26$ \\ 
A010    &K4Ve            &$  4540$&   $   $ 407 & h & $     3.08   $ &$   42$&$    *$ \\ 
A013    &M4e             &$  3289$&   $ < $  60 & l & $ <  -0.67   $ &$    *$&$    *$ \\ 
A028    &G9V             &$  5230$&   $   $  39 & h & $     1.58   $ &$   22$&$    *$ \\ 
A030    &M5Ve            &$  3170$&   $ < $  60 & l & $ <  -0.79   $ &$    *$&$    *$ \\
\end{tabular}
\end{table*}

\addtocounter{table}{-1}
\begin{table*}
\tabcolsep=5pt
\centering
\caption[]{continued.
} 
\begin{tabular}{llllllrl} 
\noalign{\smallskip}     
\hline 
\hline 
\noalign{\smallskip}    
field & Sp. type  &$T_{\rm eff}$  &$W$(\LiI ) &  & 
$\log N$(Li) 
&\multicolumn{1}{l}{$v_{\rm rot,1}$}&$v_{\rm rot,2}$\\
      &           &  [K]         &  [m\AA ]  &  &                & [\kms ] &       [\kms ] \\     
\noalign{\smallskip}    
\hline 
\noalign{\smallskip}    
B004    &G5II            &$  4957$&   $   $  33 & h & $     1.19   $ &$<   5$&$    *$ \\ 
B008    &G5              &$  5553$&   $   $ 121 & h & $     2.58   $ &$    9$&$   18$ \\ 
B013    &K2Ve            &$  4836$&   $   $  30 & h & $     0.99   $ &$   31$&$    *$ \\ 
B018    &K5Ve            &$  4406$&   $   $  36 & h & $     0.50   $ &$   12$&$    *$ \\ 
B018    &K9              &$  3936$&   $   $  64 & l & $     0.16   $ &$    *$&$    *$ \\ 
B023    &K7e             &$  4150$&   $ < $  60 & l & $ <   0.42   $ &$    *$&$    *$ \\
\end{tabular}
\end{table*}

\addtocounter{table}{-1}
\begin{table*}
\tabcolsep=5pt
\centering
\caption[]{continued.
} 
\begin{tabular}{llllllrl} 
\noalign{\smallskip}     
\hline 
\hline 
\noalign{\smallskip}    
field & Sp. type  &$T_{\rm eff}$  &$W$(\LiI ) &  & 
$\log N$(Li) 
&\multicolumn{1}{l}{$v_{\rm rot,1}$}&$v_{\rm rot,2}$\\
      &           &  [K]         &  [m\AA ]  &  &                & [\kms ] &       [\kms ] \\     
\noalign{\smallskip}    
\hline 
\noalign{\smallskip}    
C095    &K1IV            &$  4624$&   $ < $  10 & h & $ <   0.20   $ &$<   5$&$    *$ \\ 
C106    &M2e             &$  3524$&   $ < $  60 & l & $ <  -0.40   $ &$    *$&$    *$ \\ 
C120    &M5e             &$  3170$&   $ < $  60 & l & $ <  -0.79   $ &$    *$&$    *$ \\ 
C125    &K0IV            &$  4775$&   $ < $  10 & h & $ <   0.41   $ &$    6$&$    *$ \\ 
C143    &K2V             &$  4836$&   $   $  75 & h & $     1.44   $ &$   18$&$    *$ \\ 
C146    &K8e             &$  4041$&   $ < $  60 & l & $ <   0.31   $ &$    *$&$    *$ \\
\end{tabular}
\end{table*}

\addtocounter{table}{-1}
\begin{table*}
\tabcolsep=5pt
\centering
\caption[]{continued.
} 
\begin{tabular}{llllllrl} 
\noalign{\smallskip}     
\hline 
\hline 
\noalign{\smallskip}    
field & Sp. type  &$T_{\rm eff}$  &$W$(\LiI ) &  & 
$\log N$(Li) 
&\multicolumn{1}{l}{$v_{\rm rot,1}$}&$v_{\rm rot,2}$\\
      &           &  [K]         &  [m\AA ]  &  &                & [\kms ] &       [\kms ] \\     
\noalign{\smallskip}    
\hline 
\noalign{\smallskip}    
E170    &M5e             &$  3170$&   $   $   * &   & $        *   $ &$    *$&$    *$ \\ 
E179    &K4IVe           &$  4207$&   $   $  66 & h & $     0.53   $ &$   18$&$    *$ \\ 
E221    &G9III           &$  4726$&   $ < $  10 & h & $ <   0.34   $ &$    8$&$    *$ \\ 
E256    &K4V             &$  4540$&   $ < $  10 & h & $ <   0.09   $ &$    6$&$    *$ \\ 
E262    &K0V-IV          &$  5152$&   $   $  31 & h & $     1.39   $ &$    6$&$    *$ \\ 
F002    &K1IV            &$  4624$&   $ < $  10 & h & $ <   0.20   $ &$   13$&$    *$ \\
\end{tabular}
\end{table*}

\end{document}